\documentclass[pdflatex,sn-nature]{sn-jnl}

\usepackage{graphicx}%
\usepackage{multirow}%
\usepackage{amsmath,amssymb,amsfonts}%
\usepackage{amsthm}%
\usepackage{mathrsfs}%
\usepackage[title]{appendix}%
\usepackage{xcolor}%
\usepackage{textcomp}%
\usepackage{pifont}
\usepackage{manyfoot}%
\usepackage{booktabs}%
\usepackage{algorithm}%
\usepackage{algorithmicx}%
\usepackage{algpseudocode}%
\usepackage{listings}%
\usepackage{hyperref}%

\UseRawInputEncoding

\def\kms{km~s$^{-1}$}

\def\lsim    {{_<\atop^{\sim}}}

\newcommand{\Lx}{\ensuremath{L_\mathrm{x}}}

\newcommand{\ergsec}{\ensuremath{\mathrm{erg\,s}^{-1}}}

\newcommand{\Msun}{\ensuremath{M_\odot}}
\newcommand{\ml}{\ensuremath{\Msun~\mathrm{yr}^{-1}}}

\newcommand{\cxo}{\textit{Chandra X-ray Observatory}}
\newcommand{\chandra}{\textit{Chandra}}

\newcommand{\hst}{\textit{HST}}
\newcommand{\spitzer}{\textit{Spitzer}}

\newcommand{\rosat}{\textit{ROSAT}}

\newcommand{\einstein}{\textit{Einstein}}
\newcommand{\ee}[2]{\ensuremath{{#1}\!\times\!10^{#2}}}

\newcommand{\aj}{Astron. J.}   
\newcommand{\apj}{Astrophys. J.}   
\newcommand{\apjl}{Astrophys. J. Lett.}   
\newcommand{\apjs}{Astrophys. J. Suppl. Ser.}   
\newcommand{\apss}{Astrophys. Space Sci.}   
\newcommand{\aap}{Astron. Astrophys.}   
\newcommand{\aapr}{Astron. Astrophys. Rev.}   
\newcommand{\mnras}{Mon. Not. R. Astron. Soc.}   
\newcommand{\nat}{Nature} 
\newcommand{\nastro}{Nat. Astron.} 
\newcommand{\prl}{Phys. Rev. Lett.}   
\newcommand{\pasj}{Publ. Astron. Soc. Jpn}   
\newcommand{\ssr}{Space Sci. Rev.}   
\newcommand{\asr}{Adv. Space Res.} 
\newcommand{\npbps}{Nucl. Phys. B Proc. Suppl.} 
\newcommand{\jastp}{ J. Atmos. Sol.-Terr. Phys.} 

\begin{document}

\title[Article Title]{25 Years of Groundbreaking Discoveries with Chandra}


\author*[1]{\fnm{Patrick} \sur{Slane}}\email{slane@cfa.harvard.edu}
\equalcont{These authors contributed equally to this work.}

\author[1]{\fnm{\'Akos} \sur{Bogd\'an}}\email{abogdan@cfa.harvard.edu}
\equalcont{These authors contributed equally to this work.}

\author[2,3]{\fnm{David} \sur{Pooley}}\email{dpooley@trinity.edu}
\equalcont{These authors contributed equally to this work.}

\affil*[1]{\orgname{Center for Astrophysics \ding{120} Harvard \& Smithsonian}, \orgaddress{\street{60 Garden St.}, \city{Cambridge}, \postcode{02138}, \state{MA}, \country{USA}}}

\affil[2]{\orgdiv{Department of Physics and Astronomy}, \orgname{Trinity University}, \city{San Antonio}, \state{TX}, \country{USA}}

\affil[3]{\orgname{Eureka Scientific, Inc.}, \city{Oakland}, \state{CA}, \country{USA}}



\abstract{
The \cxo\ is a mainstay of modern observational
astrophysics. With the highest angular resolution of any X-ray
facility, its imaging and spectral capabilities in the 0.5-10~keV
band have led to both unique and complementary breakthroughs in
nearly all areas of the field. Now more than a quarter century into
its mission, \chandra\ continues to provide unique information on the
contributions of compact objects to the evolution of galaxies, the
nature of supernova explosions, the impact of energetic jets from
supermassive black holes on their host environments, and the fate
of exoplanet atmospheres in systems rich with stellar flares. Here
we provide a summary of \chandra\ results -- one that is embarrassingly
incomplete, but representative of both the exquisite past and
promising future for \chandra's contributions to high energy
astrophysics and all of mainstream astronomy.
}




\maketitle

\section{Introduction}\label{sec1}
As recently as 50 years ago -- an eye blink in the history of astronomy
-- our knowledge of the high energy universe was confined to a small
catalog of discrete sources accompanied by apparently diffuse X-ray emission
filling the sky.
Today, this hazy X-ray background has been resolved into a speckled
array dominated primarily by black holes, and we recognize a sky
teeming with X-ray emission produced by objects from within our own
Solar System to the very edge of the known Universe. This revolution
has been brought about, largely, through the development of telescopes
with mirrors capable of producing images of the X-ray sky. 

The \cxo\ (\chandra) represents the current pinnacle in
such facilities. With mirrors capable of resolving sources separated
by less than $1^{\prime\prime}$, the observatory can
distinguish discrete sources from diffuse emission, image extremely small
structures, and measure proper motions and expansions of objects
over time periods of years to decades. 
With a typical total background of ~2 counts per million second
integration within a resolution element of $\lsim 1^{\prime\prime}$ radius, \chandra\ is able to detect
exceedingly faint objects -- both very low luminosity nearby sources
and luminous sources at high redshift.
These capabilities open
unique avenues of study that have resolved the diffuse X-ray
background into point sources, separated emission from discrete
sources in external galaxies to reveal the intervening interstellar
medium (ISM), identified neutron stars (NSs) and clumps of metal-rich ejecta
in supernova remnants (SNRs), and resolved shocks in merging galaxy clusters.

\chandra's high resolution mirrors are accompanied by two detectors, each capable of detecting individual X-ray photons -- a CCD
camera that offers both imaging and moderate spectral resolution spectroscopy, and
a microchannel device that provides slightly higher resolution imaging
along with fast timing capabilities. Each camera has both imaging
and spectroscopy arrays, the latter used to read out X-rays dispersed
through one of two transmission grating spectrometers that can be inserted into the
beam to provide spectra with a resolving power of $\sim$100--1000 over
the 0.25--10 keV bandpass. This powerful combination has been used
to push the forefront of high energy astrophysics, breaking completely
new ground and establishing X-ray observations as a crucial component
of mainstream astronomical investigations. 

Here we present an overview of highlights from the first quarter
century of studies with \chandra. A complete summary is far beyond
the scope of this article, but the reader is referred to \cite{2019cxro.book.....W}
for a more expansive treatment. This
overview is organized around contributions to three fundamental
questions of modern astrophysics: How Did We Get Here? How Does the Universe Work?, and  Are We Alone?

\section{Key Insights on Our Origins}

Throughout a star's birth, life, and death, whether on its own or with a partner in a binary, there are key high-energy processes that provide unique information on how the stars, planets, and elements in our Galaxy came to exist in their present form and distribution.  Stars, their explosions, and the explosions of their remnants, are the engines by which a galaxy is populated with a rich assortment of heavy elements.  The contributions of previous generations of stars can be estimated by the population of compact objects they have left behind: white dwarfs, neutron stars, and black holes. The latter two are most easily observed by their production of X-rays in accreting binaries.

\subsection{Stellar Birth and Growth}
Many key insights can only be learned studying stars in the dense regions in
which they form, for which \chandra's spatial resolution and ability to
detect highly embedded sources is essential.  Young stars of all masses emit
X-rays, with the X-ray luminosity decreasing with age ($t$) as $t^{-1/2}$, which is driven by the convective turnover rate \cite{2003SSRv..108..577F}; M stars, for example, can have elevated X-ray emission for up to 5 billion years.  \chandra\ observations of young stellar clusters can identify up to twice as many young stars as those identified by an IR-excess using \spitzer.  The larger samples made possible by \chandra\ observations have led to key discoveries on the formation, structure, and populations of young stellar clusters and subclusters.  For example, NGC 2024 and the Orion Nebula Cluster were found to have older populations in the outskirts of the clusters and younger populations in the cores, implying that cluster formation is a slow process, contrary to standard, simple models of star formation \cite{2014ApJ...787..109G}.  In a detailed survey of 17 star-forming regions, four distinct classes of spatial structure --- long chains of subclusters, clumpy structures, isolated clusters with a coreâhalo structure, and isolated clusters well fit by a single isothermal ellipsoid --- were identified \cite{2014ApJ...787..107K}, and further support was found for a universal X-ray luminosity function \cite{2015ApJ...802...60K}, which is closely related to the initial mass function.

\begin{figure}[h]
\centering
\includegraphics[width=1.0\textwidth]{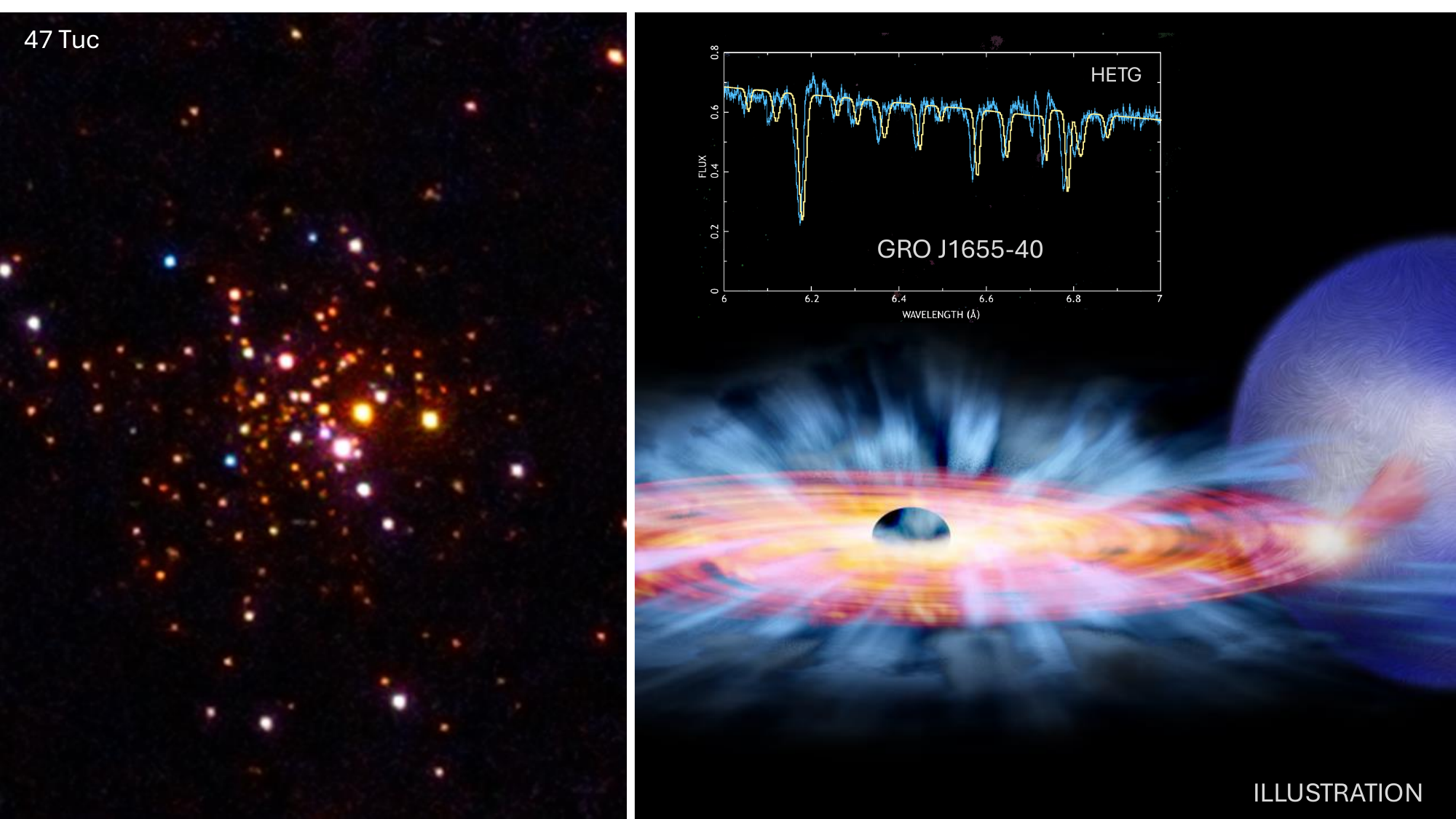}
\caption{{\bf \chandra\ studies of stellar systems and X-ray binaries.}
{\it Left:} \chandra\ image (0.5--1.5 keV in red, 1.5--2.5 keV in green,
4.0--6.0 keV in blue) of the inner $1.8' \times 2.3'$ of the globular cluster
47 Tuc reveals a large and diverse population of close X-ray binaries and
their progeny, including low-mass X-ray binaries, millisecond pulsars,
cataclysmic variables, and magnetically active BY Dra binaries (credit:
NASA/CXC/Michigan State/A.~Steiner et al. \cite{2013ApJ...765L...5S}. {\it
Right:} Artist's impression of the wind coming off the accretion disk in a
black hole binary (credit: NASA/CXC/M. Weiss).  The inset shows a portion of
the measured \chandra\ absorption spectrum of GRO~J1655$-$40 in blue and a
model in yellow at the expected wavelengths without bulk outflow. Credit:
NASA/CXC/U.\ Michigan/J.~Miller et al. \cite{2008ApJ...680.1359M}}\label{starsfig}
\end{figure}

The  growth of pre-main sequence stars via steady accretion can be probed in
X-rays produced by the shocks formed by the accretion stream being
magnetically funneled onto the stellar surface. \chandra\ High Energy
Transmission Grating (HETG) observations of TW Hya show how the X-ray line
ratios allow determination of the mass infall rate and details of the coronal
emission\cite{2010ApJ...710.1835B}.  The \chandra\ high resolution spectra
revealed the presence of both a $\sim 10^7$K coronal plasma and a $\sim 2.5
\times 10^6$K plasma due to the accretion shock.    The electron density determined from He-like Ne IX agreed with accretion models, but the density determined from O VII is a factor of 4 lower, a finding which necessitated the development of a new model of an accretion-fed corona.

Extreme episodic accretion, such as during FU Orionis (FUor) type outbursts \cite{2014prpl.conf..387A} have the highest stellar accretion rates. X-ray and infrared observations can probe these heavily obscured sources, and again \chandra's resolution is critical to associate any X-ray emission in the direction of an outburst with the actual star undergoing an FUor outburst; X-ray observations with poorer spatial resolution can be ambiguous because crowding can be an issue \cite[e.g.,][]{2014A&A...570L..11L,2020AJ....159..221S}.  For example, the namesake of the class, FU Ori itself, was shown by \chandra\ to have confusion of its X-ray emission with a companion $0.20''$ away \cite{2010ApJ...722.1654S}.  Such resolution was also necessary for the direct X-ray detection of a protostar outflow \cite{2001Natur.413..708P}, only possible with \chandra, which detected shock-heated material at the leading edge of the outflow at a temperature of $\sim$$10^6$ K.

\subsection{Stellar Life}
The production of X-rays by pre-main sequence and young stars has direct and
potentially devastating consequences on both the survival of the
proto-planetary disks at early phases and the stripping and chemistry of
planetary atmospheres at later phases.  The high energy radiation of massive
stars can accelerate the photo-evaporation of proto-planetary disks, causing
them to dissipate and eventually disappear faster than normal; \chandra\ has
indeed shown that in regions of Cygnus OB2 with lower stellar densities and
fewer high mass stars, the fraction of young stars with disks is $\sim$40\%,
whereas in higher density regions the fraction is $\sim$18\%
\cite{2023ApJS..269...13G}.  Later in a star's life, its stellar activity can
have a large impact on its planets' habitability, and coronal mass ejections
in particular are a severe habitability concern \cite{2010AsBio..10..751S},
yet their rate of occurrence is not well constrained.  \chandra\ has detected
one extreme event (mass of $\sim$\ee{1.2}{21}\,g and kinetic energy of
$\sim$\ee{5}{34}\,erg) with HETG-measured Doppler shifts of 100--400\,\kms in
S XVI, Si XIV, and Mg XII lines \cite{2019NatAs...3..742A}; however,
substantial numbers of coronoal mass ejection characterizations in X-rays will need to wait for future X-ray missions like Lynx.

\subsection{Stellar Death}
At the end of a massive star's life, the cessation of nuclear fusion in the core begins the process of core collapse and subsequent supernova (SN) explosion.  As the cast off stellar ejecta interact with the surrounding medium --- itself formed from the progenitor star through its winds --- both are heated and emit continuum and line emission in X-rays.  In the weeks, months, and years after explosion, these X-rays can probe the late stage evolution of the massive star, revealing hundreds to thousands of years of pre-supernova mass loss history, information impossible to gather in real time.  

In general, the high X-ray luminosities of all types of core-collapse SNe
dominate the total radiative output of the SNe starting at an age of about
one year. As the SN outgoing shock emerges from the star, its characteristic
velocity is $\sim$10$^4$\,\kms, and the density distribution in the outer
parts of the ejecta can be approximated by a power-law in radius,
$\rho\propto r^{-n}$, with $7\lesssim n \lesssim 20$. The outgoing shock
propagates into a dense circumstellar medium (CSM) formed by the pre-SN
stellar wind, whose density follows $\rho=\dot{M}/4\pi r^2v_w$ where
$\dot{M}$ is the pre-SN mass-loss rate and $v_w$ is the pre-SN stellar wind
velocity. The collision between the SN ejecta and CSM also produces a
``reverse'' shock, which travels outward at $\sim$10$^3$ \kms\ slower than
the fastest ejecta. Interaction between the outgoing shock and the CSM
produces a hot shell ($\sim$10$^9$ K), while the reverse shock produces a
denser, cooler shell ($\sim$10$^7$ K) with much higher volume emission measure from which most of the observable X-ray emission arises.  By following the evolution of the X-ray (along with radio and optical/UV) emission, \chandra\ observations can measure the temperature and radiative cooling rate and determine the structure of the SN ejecta, structure of the CSM, presence or absence of a dense shell of cold gas surrounding the X-ray emitting region, and details of the star's pre-SN evolution.

Because these investigations happen in nearby galaxies (tens of Mpc), which have their own heterogeneous populations of X-ray sources, \chandra's sub-arcsecond resolution is required to give an accurate picture of the X-ray emission from young SNe, and it has constrained the steady and episodic mass loss histories of a number of SN progenitors.  For example \chandra\ observations have enabled the determination of steady mass loss rates of $\dot{M}\approx (1-2) \times 10^{-4}$~\ml\ from the progenitor of the Type IIn SN 1998S, $\dot{M}\approx 2 \times 10^{-6}$~\ml\ from the progenitors of the Type II-P SNe 1999em and 2004et, and $\dot{M}\approx 3 \times 10^{-4}$~\ml\ from the progenitor of the Type IIb SN 2003bg \cite{2002ApJ...572..932P, 2006ApJ...651.1005S, 2007MNRAS.381..280M}.  More complicated mass-loss histories involving the stripping of the outer envelopes of massive stars have been revealed by \chandra\ observations of SNe 2004dk and 2014C \cite[e.g.,][]{2022ApJ...939..105B, 2022ApJ...930...57T, 2018MNRAS.478.5050M, 2019ApJ...883..120P}.

\chandra\ has also detected heavy elements in prompt X-ray emission months to years after SN explosion with clear evidence for overabundances of Ne, Al, Si, S, Ar, and Fe in SN 1998S \cite{2002ApJ...572..932P} and  indications of overabundances of O, Ne, and Ni in SN 2003bg \cite{2006ApJ...651.1005S}.  \chandra\ images of the longer lived supernova remnants (SNRs), both those resulting from core collapse of a massive star and those resulting from thermonuclear explosions of white dwarfs, over the past 25 years show the direct and dramatic expansion of heavy elements into interstellar space, beautifully confirming this key phase in the life cycle of stars.   Additionally, these SNRs provide rich and unparalleled laboratories for exploring fundamental shock physics, cosmic ray acceleration, and a number of other topics \S\ref{physics:snr}.

\subsection{Stellar Remnants}
The neutron stars and black holes left behind by core-collapse SNe are among the most exotic objects in the universe.  When one of these compact objects is in a binary and accretes matter from a companion star, the X-ray emission can give unique insights into not only fundamental physics (e.g., accretion, General Relativity, and quantum mechanics)
but also fundamental astrophysics (e.g., star formation history, stellar evolution, and compact object merger).  As an example, \chandra\ HETG observations of the black hole binary GRO J1655$-$40 (Figure \ref{starsfig}, right) revealed a rich spectrum with more than 100 blue-shifted absorption lines, primarily from H- and He-like ions of O, Si, Mg, and Fe, indicating the presence of a wind blowing from the black hole accretion disk \cite{2008ApJ...680.1359M}; detailed modeling shows the wind cannot be driven by radiation or photoionization, but could be due to heating by magnetic turbulent viscosity in the disk.

In addition, these X-ray binaries, as a population, serve as one of the best
observational probes of binary stellar evolution
\cite[e.g.][]{2006AdSpR..38.2937F}.  With its sensitivity and spatial
resolution, \chandra\ observations have established a strong correlation
between a galaxy's high-mass X-ray binaries and its star-formation rate
\cite[e.g.,][and references therein]{2012MNRAS.419.2095M}, with the
collective X-ray luminosity of a galaxy's high-mass X-ray binaries related to
its star-formation rate (SFR) as $L_\mathrm{X}^\mathrm{HMXB}\,(\ergsec) =
\ee{2.61}{39}\ \mathrm{SFR\,(\ml)}$. \chandra\ observations of 30 early-type
galaxies \cite{2011ApJ...729...12B} show a scaling between a galaxy's
low-mass X-ray binaries and its stellar mass measured by $K$-band light as
$L_\mathrm{X}^\mathrm{LMXB}\,(\ergsec) = 10^{29} L_K\,(L_{K\odot})$ where
$L_{K\odot}$ is the K-band luminosity of the Sun. 

These results, along with the populations synthesis models they inform, indicate that X-ray binaries will produce the dominant X-ray emission of the early ($z \gtrsim 6$) Universe \cite{2016ApJ...825....7L, 2017ApJ...840...39M} and be the main heat source of the intergalactic medium prior to the epoch of reionization \cite[e.g.,][]{2014MNRAS.443..678P}.

\subsection{X-ray Binary Formation}
Understanding the formation of X-ray binaries is one the most difficult problems in astrophysics, due in large part to the complexities of common envelope evolution \cite[e.g.][]{2013A&ARv..21...59I}.  There is one scenario, however, in which \chandra\ has enabled substantial progress, clearing up several decades-old questions.

Since the discovery of luminous ($\gtrsim 10^{36}\,\ergsec$) low-mass X-ray binaries in the 1970s, it was noted that their formation rate per unit mass  is orders of magnitude higher in globular clusters than in the Galactic disk \cite{1975ApJ...199L.143C, 1975Natur.253..698K},   An additional population of low-luminosity ($\Lx \lesssim 10^{35}~\ergsec$) globular cluster X-ray sources was discovered with the \einstein\ satellite \citep{1983ApJ...267L..83H, 1983ApJ...275..105H} and further explored with \rosat\ \cite{2001A&A...368..137V}.  

However, the nature of this low-luminosity population remained elusive for nearly two decades. Early, deep \chandra\ observations \cite[e.g.][]{2001Sci...292.2290G, 2001ApJ...563L..53G, 2002ApJ...569..405P, 2002ApJ...578..405R, 2002ApJ...573..184P}  not only discovered many more such sources (Figure \ref{starsfig} left) but also, where \hst\ observations were available, allowed for initial progress in classifying the members of this low-\Lx\ population as either quiescent low-mass X-ray binaries, cataclysmic variables, millisecond pulsars, or chromospherically active main-sequence binaries.  Although a heterogeneous mix, all are close binary systems or their progeny.  

\chandra\ observations established that the number of X-ray sources in a globular cluster scales with the encounter frequency significantly better than with the mass of the cluster \citep{2003ApJ...591L.131P}, showing that these X-ray sources are largely dynamically formed.  \chandra\ also found that a  population dominated by quiescent low-mass X-ray binaries had a very strong dependence on encounter frequency \cite{2003ApJ...591L.131P, 2003ApJ...598..501H, 2003A&A...400..521G} while a population dominated by cataclysmic variables had a weaker dependence \cite{2006ApJ...646L.143P}, suggesting the cataclysmic variable population has both primordial and dynamical contributions. 

\subsection{Compact Object Mergers}
The gravitational wave event GW170817 resulted from the merger of two neutron
stars and was accompanied by electromagnetic emission across the spectrum,
from $\gamma$-rays to radio.  Because of \chandra's sensitivity, it detected
GW170817 in X-rays before other satellites
\cite[e.g.,][]{2017ApJ...848L..25H, 2017ApJ...848L..20M, 2017Natur.551...71T}
and can continue to follow its evolution \cite{2020MNRAS.498.5643T}.  The X-ray emission to date is well described by a highly relativistic jet launched by the merger remnant and interacting with the surrounding medium. 

In addition to synthesizing large amounts of r-process elements, the event is notable because the merged object has a mass of 2.74 \Msun\ \cite{2017PhRvL.119p1101A} and could be either a black hole or massive neutron star. Long term X-ray monitoring with \chandra\ could reveal X-ray emission associated with a spinning neutron star \cite[e.g.,][]{2019MNRAS.483.1912P}, whereas a black hole would not be expected to produce substantial late-time X-ray emission.

\section{An Unprecedented View of the Physical Workings of the Universe}

\noindent
Our understanding of the Universe is built upon the assumption that the principles of physics as we known them apply everywhere. Yet the vast extremes and complexities encountered in environments from NSs and BHs to the very formation of structure in the early stages after the Big Bang present challenges in unfolding the structure of the known Universe and its contents. Observations of high energy phenomena provide unique insights, and \chandra's contributions in these areas have been immense.

\subsection{Supernova Remnants}
\label{physics:snr}
SN explosions and their remnants play a crucial role in shaping the dynamics and chemical evolution of galaxies, enriching their environments with metals, driving shocks that heat the ISM and trigger new star formation, and acting as particle accelerators to produce an energetic population of cosmic-rays. NSs, formed in core-collapse (CC) SNe, represent the most extreme density and magnetic field conditions in the observable universe, and their properties contain signatures of the explosive process in which they were formed.  

\begin{figure}[h]
\centering
\includegraphics[width=1.0\textwidth]{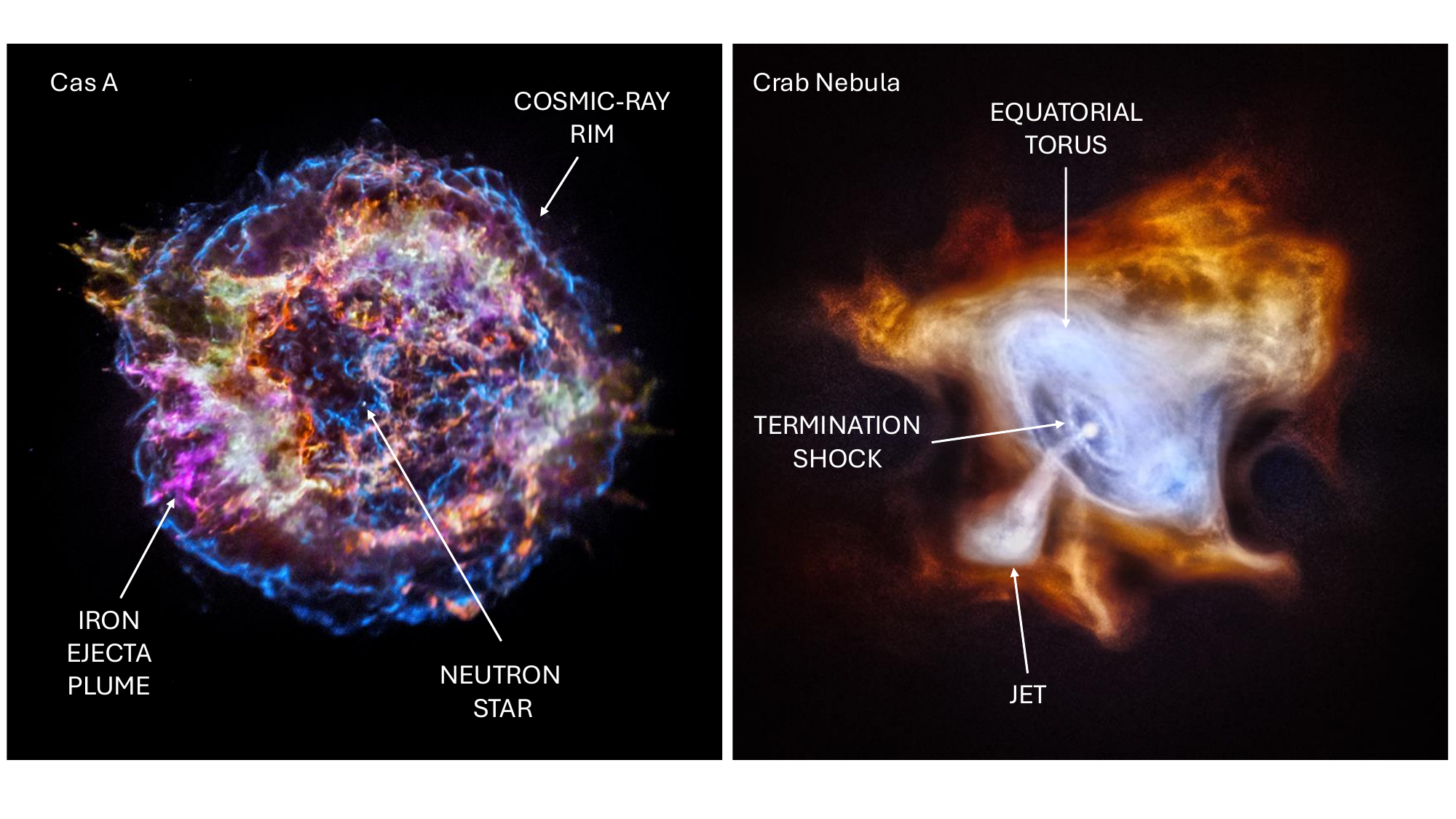}
\vspace{-0.2in}
\caption{{\bf \chandra\ details on the aftermath of stellar explosions.}
Left: \chandra\ image of Cas~A (red: 0.5-1.5keV; green 1.5-2.5 keV, blue: 4.0-6.0 keV). The young NS can be seen in the remnant center. Shocked ejecta in the interior is distributed in a complex of clumps and filamentary features. The distinct magenta-colored emission in the southeast is dominated by Fe-L emission (a broad band 
around 1 keV) from Fe-rich material while a shell of synchrotron emission (blue) identifies sites of particle acceleration to cosmic ray energies. [Credit: NASA/CXC/SAO] Right: \chandra\ image of Crab Nebula (red=0.5-1.2 keV, green=1.2-2.0 keV, blue=2.0-7.0 keV). The central pulsar is surrounded by an inner ring marking the wind termination shock and by an equatorial torus. A prominent jet identifies the direction of the pulsar rotation axis. The emission is hard (blue) near the pulsar, but softer (red) in the outskirts of the nebula due to synchrotron losses from the most energetic particles. [Credit: NASA/CXC/SAO] }\label{snrfig}
\end{figure}

Because CC SNRs can result from a broad range of progenitor masses and final stellar configurations, connecting the observed properties to progenitor type provides critical information on the details of these explosive events. Cas~A  provides an illustrative example of the numerous breakthroughs in \chandra\ SNR studies. The image (Figure 2, left) reveals a complex network of shock-heated ejecta surrounded by a thin rim of synchrotron emission from electrons with energies approaching 100~TeV. Spatially-resolved spectroscopy establishes a clear overturn of the ejecta \cite{2000ApJ...528L.109H}, resulting in Fe being mixed to the outermost regions of the remnant despite being formed near the core.  A census of the ejecta composition and mass, from spectra resolved on small spatial scales, has identified regions where Fe is accompanied by other products of incomplete Si burning, while others are nearly pure Fe produced during complete Si burning \cite{2012ApJ...746..130H}.  The total inferred mass is $\sim 3.5 M_\odot$, of which $\sim 0.1 M_\odot$ is Fe. However, because Cas~A is young, the reverse shock has not yet propagated through (and heated) all of the ejecta. Observations with JWST uncover unshocked ejecta in the central regions, along with a wealth of complementary information about Cas~A \cite{2024ApJ...965L..27M}.

The very first \chandra\ observations of Cas~A identified a faint point source in the remnant center -- the NS formed in the SN explosion. Its properties, however, differ wildly from those expected at the time. For example, the Crab Pulsar and its nebula (Figure 2, right), if moved to the same distance, would outshine the Cas~A NS by a factor of $\sim 10^4$. No pulsations are detected from the Cas~A NS, nor is there any evidence of an associated wind nebula (see \S3.2); the emission is completely derived from cooling of the hot NS interior, placing constraints on the equation of state of matter at the extreme central densities \cite{2023MNRAS.518.2775S}. Additional \chandra\ discoveries of such so-called ``compact central objects'' have now been identified in other young SNRs as well.

Early observations of SN~1987A -- barely resolved by \chandra, with a
diameter of only $\sim 1.2^{\prime\prime}$ -- showed the blast wave beginning
to interact with dense clumps at the inner edge of an equatorial ring,
producing bright spots also observed in optical observations, and eventually
entering the bulk of the ring material, resulting in an increase in the X-ray brightness, and an accompanying decrease in the observed expansion rate \cite{2016ApJ...829...40F}. Most recently, \chandra\ observations show changes in the electron temperature and volume emission measure that indicate that SN 1987A is expanding out of the equatorial ring and into a new region of CSM, and also reveal the presence of an Fe K line that may represent onset of the reverse shock interacting with SN ejecta \cite{2024ApJ...966..147R}. Observations of SN~1987A, as it transitions from an SN to an SNR, continue as a legacy of \chandra.

X-ray studies of SNRs from Type Ia SNe -- known to originate from the
complete disruption of a white dwarf star in a binary system, and used widely as
cosmological yardsticks -- provide signatures of their evolutionary paths, in
particular on whether the progenitor systems contain a single white dwarf accreting
from a normal stellar companion, or two white dearfs merging through orbital losses
from gravitational wave emission (the so-called ``single degenerate'' and
``double degenerate'' evolutionary scenarios). \chandra\ observations of
Kepler's SNR (the remnant of SN~1604) classify this as a Type Ia remnant
based on the dominant Si, S, and Fe emission, and the remnant displays
substantial shocked CSM in the north, suggesting interaction with mass lost
from a fairly massive progenitor -- either that of the white dwarf or a companion star. This may provide evidence for single-degenerate progenitor for this system \cite{2007ApJ...668L.135R}.

Additional approaches have been used to classify SNR types based on \chandra\ observations.
Moment analysis of SNR morphologies, for example, shows that CC remnants display a
lower degree of spherical symmetry and mirror symmetry than those
from Type Ia explosions -- presumably associated with evolution in
complex environments near their birth sites, and/or asymmetries in
the explosions themselves \cite{2011ApJ...732..114L}. Studies of the Fe K ionization
state also identify a distinction between CC and Type Ia
remnants, with the former displaying more highly ionized states due
to the higher density of the environments \cite{2014ApJ...785L..27Y}. These new methods of differentiating between CC and Ia origins, even for middle-aged SNRs, offer an important tool for SNR demographics.

\subsection{Particle Acceleration}

With rapid shocks and large kinetic energy reserves, SNRs have long been
recognized as potential sites of cosmic-ray acceleration. The
identification of nonthermal X-rays from SNR shell clearly demonstrates
the presence of electrons with energies approaching energies of
$\sim 10^{14}{\rm\ eV}$.  All historical SNRs studied with \chandra\
reveal such emission components, most with this nonthermal emission
confined to narrow rims (e.g., Figure 2, left). The rim thickness
is limited by synchrotron losses in the diffusion and/or advection
of relativistic particles downstream of the shock, and provides
important information on the magnetic field properties in the
acceleration region \cite{2011Ap&SS.336..257R, 2012A&ARv..20...49V}. 
These radiative losses also limit the maximum energies of the
accelerated electrons, providing a connection between the spectral
cutoff, the shock velocity, and the mean free path of the particle
transport.

The high resolution offered by \chandra\ provides for measurements of the
shock thickness and the expansion rate over multi-epoch
observations that address the particle acceleration properties of
different SNRs. Combined with X-ray polarization measurements with
the Imaging X-ray Polarimetry Explorer (IXPE), these observation establish constraints on the turbulence
levels and size scales in the particle acceleration regions
\cite{2024Galax..12...59S}.

\subsection{Neutron Star Kicks}
Many pulsars are known to have large velocities, presumably imparted in the supernova explosion process. The mechanism by which such ``kicks'' are produced is poorly understood, as are many details of the explosions themselves. Anisotropic neutrino emission can produce NS kicks \cite{2006ApJS..163..335F}, and are expected to result in alignment between the spin axis and the kick velocity \cite{2001ApJ...549.1111L} due to the long duration of the neutrino process relative to the initial spin period of the proto-NS \cite{1998Natur.393..139S}. Alternatively, convective instabilities can result in asymmetric ejection of the supernova ejecta, with momentum-conserving recoil of the NS in the opposite direction \cite{2006A&A...457..963S, 2013A&A...552A.126W}.  \chandra\ observations of young NSs in SNRs yield important constraints on this process. Proper motion measurements from observations taken over large time baselines have established the kick velocity directions (in the plane of the sky). Comparison of the proper motion direction with the distribution of ejecta material measured in \chandra\ observations of these and other SNRs shows a clear  anti-correlation that supports the convective instability scenario \cite{2017ApJ...844...84H}.

\subsection{Pulsar Wind Nebulae}
\chandra\ observations of the pulsar wind nebulae (PWNe) produced by
some young NSs reveal distinct jet-torus structures. This is seen most dramatically
for the Crab Nebula (Figure \ref{snrfig}, right;
\cite{2000ApJ...536L..81W}), where the central pulsar is surrounded
by a distinct ring at which the
cold equatorial electron-positron wind from the inner underluminous region terminates in a shock as it enters the
more slowly expanding nebula. The subsequent flow is dominated by
an equatorial torus and accompanied by jets aligned with the pulsar
rotation axis. Synchrotron cooling of the energetic particles as they
diffuse to the outskirts of the nebula is seen as a steeping of the
X-ray spectrum, evident as a color change in Figure 2 (right).

Similar jet-torus structures are identified around many young NSs
\cite{2004ApJ...616..403S, 2004ApJ...601..479N, 2008AIPC..983..171K}
-- observable only with the high resolution capabilities
of \chandra\ in many cases. When observed within their host SNRs,
these PWNe provide important insights on the formation and evolution
of these systems. In G292.0+1.8, for example, the PWN jet-torus
structure \cite{2007ApJ...670L.121P} reveals a spin axis that is
oriented nearly perpendicular to the proper motion of the pulsar,
arguing against the spin-kick alignment predicted by anisotropic
neutrino emission models. In Kes~75, for which the central pulsar
shows magnetar-like properties, {\sl Herschel} emission from the innermost
unshocked ejecta -- just outside the \chandra\ boundaries of the PWN -- 
establish abundances
that, combined with hydrodynamical simulations of the PWN/SNR system,
provide evidence for both a low energy explosion and a low progenitor
mass, which are associated with explosions that give rise to magnetars
\cite{2019ApJ...878L..19T}.

Upon exiting their host SNRs, PWNe form bowshock nebulae as they
traverse the ISM. Particles eventually escape downstream to contribute
to the Galactic $e^+-e^-$ cosmic-ray population. Some such systems
display connections with long nonthermal filaments that extend in
directions unconnected with the pulsar motion or rotation axis, and appear to
represent the escape of very energetic particles from beyond the
bowshock \cite{2008A&A...490L...3B}.

\begin{figure}[t]
\centering
\includegraphics[width=1.0\textwidth]{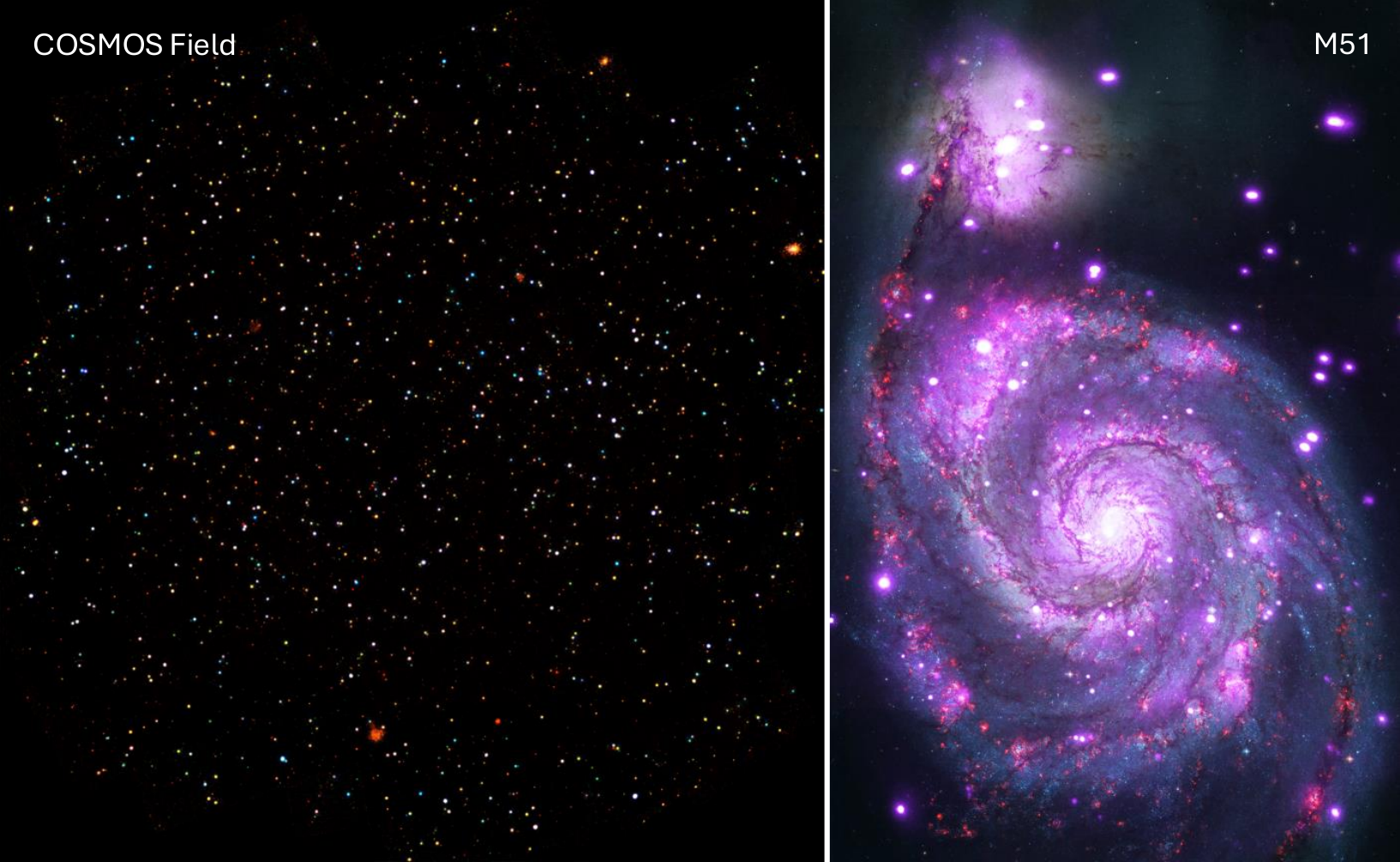}
\caption{{\bf From deep fields to nearby galaxies, \textit{Chandra}
resolves the X-ray sky into discrete sources while separating from
the truly diffuse galactic emission.} \textit{Left:} Deep X-ray image
of the Cosmic Evolution Survey (COSMOS)-Legacy Field, in which $\approx90\%$ of the CXB is
resolved into individual, faint AGN. [Credit: NASA/CXC/ICE/M.~Mezcua
et al. \cite{2018MNRAS.478.2576M}] \textit{Right:} Composite \chandra\
(pink) and Hubble Space Telescope image of the grand-design spiral
galaxy M\,51, showing diffuse X-ray-emitting hot gas along the
spiral arms and numerous point-like high-mass X-ray binaries
associated with the star-forming regions of the galaxy. [Credit:
Credit: X-ray: NASA/CXC/SAO; UV: NASA/JPL-Caltech; Optical: NASA/STScI;
IR: NASA/JPL-Caltech]}\label{fig:deepfieldfig}
\end{figure}

\subsection{Galaxies}

One of \chandra's earliest achievements was to resolve the long-standing
mystery of the cosmic X-ray background (CXB), which was seen as a diffuse
X-ray glow across the sky. Deep surveys, such as the \chandra\ Deep Field
South \cite{2017ApJS..228....2L}, successfully revealed that this glow is
composed of myriad point sources and not truly diffuse emission
(Figure~\ref{fig:deepfieldfig} left). In fact, the ultra-deep \chandra\
exposures resolved $80-90\%$ of the CXB emission into discrete sources
\cite{2006ApJ...645...95H}. Almost all of the resolved CXB intensity can be
attributed to accreting supermassive black holes (SMBHs) in distant galaxies
\cite{2007A&A...463...79G}. This census directly traces the cosmic growth of
SMBHs \cite{2008ApJ...679..118S} and shows that a large fraction of black
hole accretion occurs in obscured active galactic nuclei (AGN) only visible in X-rays \cite{2004NuPhS.132...86H}. Deep Chandra surveys have mapped the evolution of the AGN luminosity function from the local universe to high redshifts, which provides a direct measure of the SMBH accretion history across cosmic time \cite{2015A&ARv..23....1B}.

\chandraâs ability to resolve faint X-ray point sources is not only important at large cosmic distances. Long before the launch of \chandra, X-ray emission from our own Milky Way was observed as a diffuse glow. It was unclear whether this emission was produced by extremely hot gas with tens of millions Kelvin temperature, or from the blending of densely populated point sources \cite{1986PASJ...38..121K}. To resolve this question, \chandra\ observed a carefully selected patch in the Milky Way. These observations revealed that most of the Milky Way's X-ray glow can be attributed to the integrated emission of millions of faint stellar sources, such as accreting white dwarfs, coronally active binary stars, and young stellar objects \cite{2007A&A...473..857R,2009Natur.458.1142R}. Each of these objects leave their own energetic imprint, which collectively builds up the so-called Galactic ridge X-ray emission. These studies also laid the foundation for studying the diffuse X-ray glow associated with other galaxies, such as our cosmic neighbor, Andromeda. The diffuse glow from Andromeda and other galaxies, at least in part, originates from the population of very faint, discrete X-ray sources, similar to the Milky Way \cite{2007A&A...473..783R,2008MNRAS.388...56B}.

The ability to resolve individual point sources in galaxies had a fundamental
consequence: it became possible to separate the emission from discrete, point
sources from the truly diffuse gaseous emission. Studying the hot,
X-ray-emitting ISM within galaxies provides insights in  to the most
energetic events in the lives of galaxies (Figure~\ref{fig:deepfieldfig}
right). The ISM is heated by the energy input from supernovae and SMBHs,
therefore its physical properties provide an imprint on the past history of
feedback processes \cite{2009ApJ...697.2030S,2015ApJ...803L..21V}. In
addition, by studying the chemical composition of the ISM, it became possible
to probe how galaxies were enriched by heavy elements. The Milky Way's own
ISM was studied by high-resolution spectroscopy with Chandraâs Low Energy
Trasmission Grating (LETG) and HETG by detecting absorption features from elements such as Oxygen and Neon \cite{2005ApJ...624..751Y,2012ApJ...756L...8G}. These spectral diagnostics studies offered insights into the physical state, ionization balance, and metal enrichment, thereby painting a picture of our own Galaxyâs gaseous ecosystem. Beyond the optically visible regions of galaxies, Chandra probed the hot, X-rayâemitting circumgalactic medium (CGM) \cite{2013ApJ...772...97B}. In NGC\,1961, Chandra detected an expansive gaseous halo with $(3-7)\times10^6$~K temperature that extends well beyond the stellar disk \cite{2011ApJ...737...22A}. The CGM serves both as repositories for metals expelled by supernovae and stellar winds and as the long-sought reservoirs of shock-heated gas. In the standard picture of galaxy formation, baryons accrete into dark matter potentials, are shockâheated to the virial temperature, then radiatively cool to fuel star-formation, which result in the formation of galactic disks \cite{1978MNRAS.183..341W,1991ApJ...379...52W}. The detection and characterization of the CGM directly confirms this framework and links cosmological gas inflow, virial shock heating, feedback enrichment, and the regulation of galaxy growth.

\begin{figure}[!h]
\centering
\includegraphics[width=0.96\textwidth]{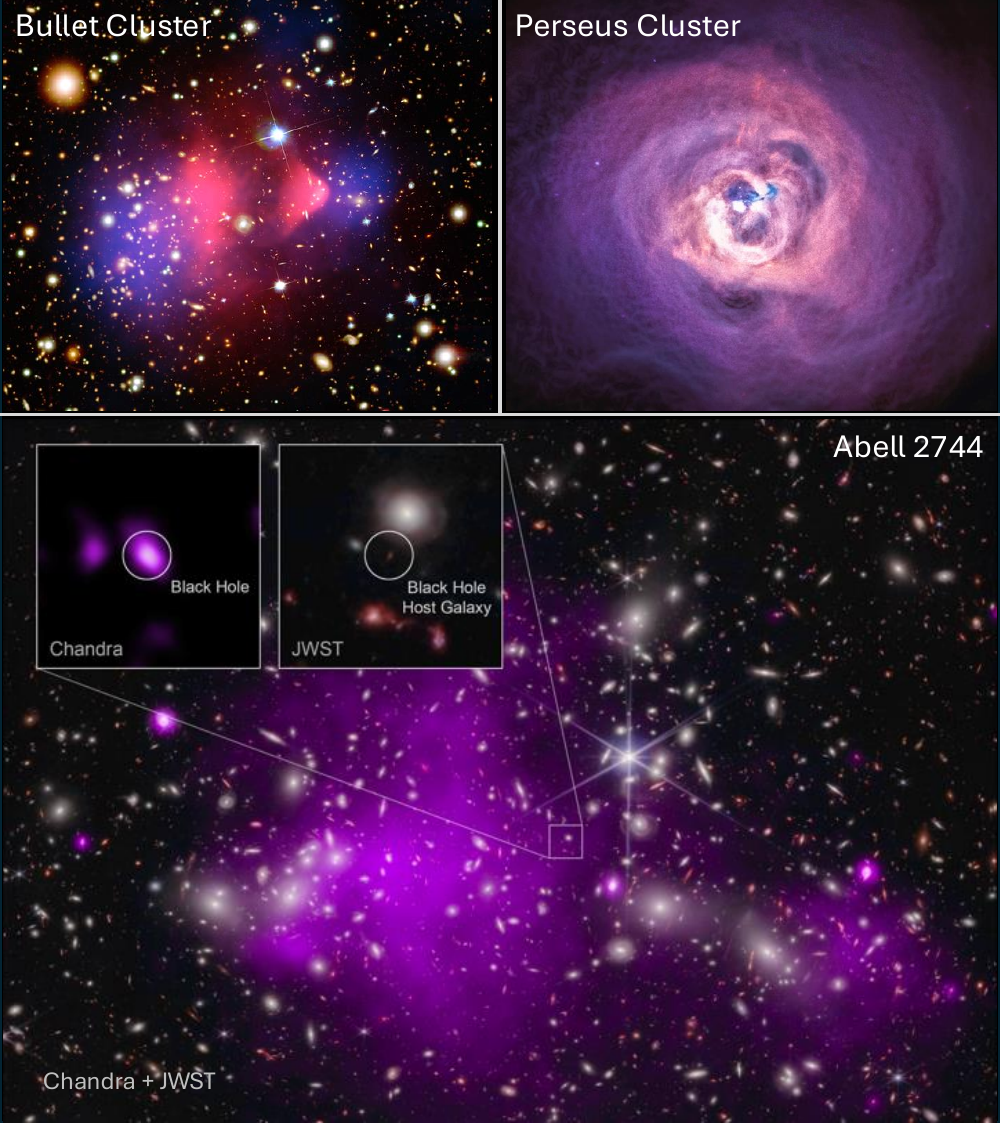}

\caption{{\bf Chandra observations of galaxy clusters transformed our
understanding of dark matter, the gas physics of the ICM, and the emergence
of the first SMBHs.}
\textit{Top left:} Composite image of Chandra X-ray (pink), optical light,
and the gravitationalâlensing mass (blue) map of the Bullet Cluster. The
offset between the  collisional ICM and the collisionless mass clumps
provides direct evidence for non-baryonic dark matter. [X-ray:
NASA/CXC/CfA/\cite{2004ApJ...606..819M}; Optical: NASA/STScI;
Magellan/U.Arizona/D.~Clowe et al. \cite{2006ApJ...648L.109C}; Lensing Map:
NASA/STScI; ESO WFI; Magellan/U.Arizona/D.~Clowe et al. \cite{2006ApJ...648L.109C}.]
\textit{Top right:} Deep Chandra view of the Perseus Cluster, revealing 
multiple AGN-inflated X-ray bubbles and ripples, illustrating AGN heating in 
a cool core.
[Credit: NASA/CXC/Univ. of Cambridge/C.~Reynolds et al. 
\cite{2020ApJ...890...59R}]
\textit{Bottom:} Chandra image of the lensing galaxy cluster Abell\,2744
(purple) overlaid on the JWST image. The inset panels show the $z=10.07$
galaxy, UHZ1. The X-ray point source is coincident with the JWST-detected
galaxy, revealing a SMBH at cosmic dawn. 
[X-ray: NASA/CXC/SAO/\'A.~Bogd\'an \cite{2024NatAs...8..126B}; Infrared: 
NASA/ESA/CSA/STScI; Image Processing: NASA/CXC/SAO/L. Frattare \& K. Arcand]}

\label{fig:clusterfig1}
\end{figure}

\subsection{Galaxy Clusters}

Galaxy clusters, containing hundreds of galaxies embedded in a massive halo of dark matter and hot gas, are excellent laboratories for both astrophysics and cosmology. While the total mass of clusters can reach $10^{15} \ \rm{M_{\odot}}$, stars confined within the cluster member galaxies account for only a few percent of this enormous mass. Instead, most of the baryonic mass is in the form of X-ray emitting hot gas with tens of millions Kelvin temperatures. This intracluster medium (ICM) accounts for $\sim10\%$ of the mass of galaxy clusters. The remaining $\sim90\%$ of mass is in the form of dark matter. Chandra observations of clusters allowed us to study the nature of dark matter, the physics of the hot ICM, and the interaction between  galaxies and their large-scale environment.

The Bullet Cluster (1E\,0657-56) is not only one of the most iconic images featuring  Chandra data, but it also provides direct empirical proof for the existence of dark matter. In this system, two galaxy clusters collided with high speeds ($\sim4500 \ \rm{km \ s^{-1}}$). Chandra mapped the distribution of the hot ICM, which contains the bulk of baryonic mass, and identified the characteristic ``bullet''-like feature (Figure~\ref{fig:clusterfig1} top left) produced by a shock front that lags behind the galaxy clusters during the collision. At the same time, gravitational lensing maps from optical data established that the distribution of dark matter is offset from the X-ray gas and coincides with the location of the galaxies \cite{2004ApJ...606..819M,2006ApJ...648L.109C}. The explanation for this offset is that during the collision between the clusters, the collisional X-ray gas slowed down and formed the shock fronts, while the non-collisional dark matter passed through unaffected, as traced by the galaxies. This provides clear evidence that an invisible and collisionless mass component, dark matter, dominates the cluster mass. This discovery not only provided direct evidence for the existence of dark matter, but also placed constraints on the self-interaction cross-section of dark matter particles, ruling out a range of exotic models \cite{2008ApJ...679.1173R}.

\begin{figure}[t]
\centering
\includegraphics[width=1.0\textwidth]{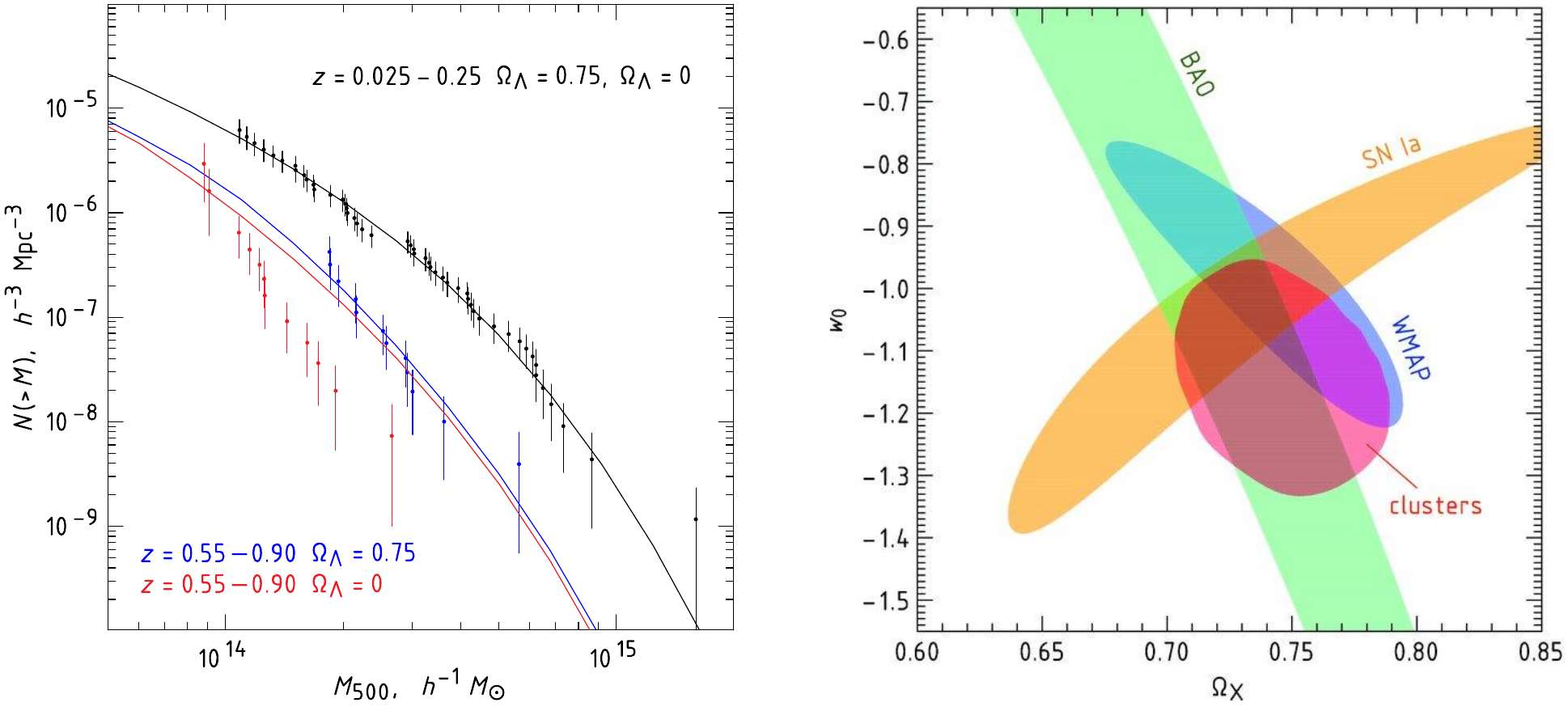}
\caption{{\bf \textit{Chandra} observations of galaxy clusters independently
confirm cosmic acceleration and  constrain the dark energy equation
of state.} \textit{Left:} Comoving cluster abundance, N$(>M)$, versus
mass, $M_{500}$, in two redshift bins, derived from \chandra\ observations.
Here $M_{500}$ is the mass within $r_{500}$, the radius enclosing 500 times
the critical density at the cluster redshift, and $h = H_0/(100 \rm\ km\
s^{-1}\ Mpc^{-1})$. Black data points show the observed abundance in the
low-redshift bin ($z = 0.025-0.25$) when cluster masses are inferred in a flat
$\Lambda$CDM cosmology. The solid black curve shows the corresponding
$\Lambda$CDM mass function. In the high-redshift bin ($z = 0.55-0.90$),
blue data points and the blue curve show the observations and the
$\Lambda$CDM prediction, while the red data points and red curve
show the same sample and prediction when masses are recomputed in a matter-
only universe. The suppression of high-mass clusters in $\Lambda$CDM
(blue/black) versus the matter-only case (red) illustrates that dark
energy slows the growth
of massive structures at early times. Error bars are $1 \sigma$.
\textit{Right:} Constraints in the $\Omega_X - w_{\rm{0}}$ plane from
galaxy clusters (red)
compared with other probes all favoring $w_{\rm{0}} \simeq -1$.
$\Omega_X = 1 - \Omega_M$ denotes the present-day dark-energy density
parameter.
Regions correspond to 68\% CL. BAO, baryon acoustic oscillations; WMAP,
Wilkinson Microwave Anisotropy Probe.
[\cite{2009ApJ...692.1060V} \copyright AAS.
Reproduced with permission.]} \label{fig:clusterfig2}

\end{figure}

Galaxy clusters are also powerful cosmological probes. Specifically, X-ray
measurements of the total and baryonic cluster mass (inferred via the
temperature and density profile of the ICM) and the evolution of cluster
number counts as a function of redshift can be used to test models of
structure growth. By tracing how the abundance of massive clusters declines
at higher redshifts, Chandra measurements confirmed the existence of dark
energy. Higher redshift clusters are expected to be less common than they
would be in a universe without the accelerated expansion of the universe
(Figure~\ref{fig:clusterfig2} left). This imprint of the dark energy was
directly probed by Chandra data of a few dozen well-characterized clusters,
and provided measurements of the cosmological matter density $\Omega_M$ and
dark energy equation-of-state parameter $w_{\rm 0}$. Specifically, the
Chandra observations of clusters demonstrated that an equation-of-state
parameter of $w\simeq-1$ provides the best-fit, which is consistent with a
cosmological constant, in agreement with the $\Lambda$ cold dark matter (CDM,
where $Lambda$ is the cosmological constant) paradigm \cite{2004MNRAS.353..457A,2009ApJ...692.1060V}. The accuracy of these cosmological parameters rival those from the cosmic microwave background and Type Ia supernovae. In addition, the degeneracy directions of cluster cosmology are orthogonal to those of other probes (Figure~\ref{fig:clusterfig2} right), providing an independent and powerful pillar of observational cosmology. 

In many cluster cores, the cooling time of the dense X-ray emitting gas is
much shorter than the age of the cluster. As a consequence, a substantial
fraction of this gas is expected to cool to low temperatures, which could
drive rapid starâformation with rates of hundreds of solar masses per year.
Initial X-ray spectra showed little to no evidence for gas at
$kT\lesssim1$~keV (where $T$ is the temperature and $k$ is the Boltzmann
constant), a discrepancy dubbed as the ``cooling-flow'' problem. High-resolution \chandra\ imaging revealed that jets from the central SMBH inflate buoyant bubbles in the ICM, thereby displacing cooling gas and injecting mechanical energy \cite{2005ApJ...635..894F}. More recently, deep \textit{XMM-Newton} Reflection Grating Spectrometer (RGS) observations have detected Fe~XVII emission from cooler gas at $0.3-0.7$~keV \cite{2024MNRAS.535.2173F}, demonstrating that a modest cooling flow, albeit with rates below the classical predictions, is present in many cool core clusters. The Perseus Cluster provides one of the most vivid examples of AGN feedback in action. \chandra\ data show two roughly symmetric, $\sim10$~kpc bubbles, where radio observations reveal relativistic plasma filling the cavities.  These bubbles are surrounded by ripple-like surface brightness fluctuations that are weak shocks propagating through the ICM \cite{2000MNRAS.318L..65F,2003MNRAS.344L..43F} (Figure~\ref{fig:clusterfig1} top right). These features indicate repeated AGN outbursts in the past that inflated bubbles and drove waves through the ICM. The energy input injected by the SMBH is sufficient to heat the gas and offset most radiative losses, and suppress star formation in cluster cores. Nevertheless, a small amount of low-temperature gas persists, and its ultimate fate is still under active investigation. Overall,  Chandra observations did not only address the classical cooling-flow problem but also demonstrated that the SMBH in the central galaxy can profoundly influence the thermal state of the entire cluster core \cite{2007MNRAS.381.1381S}. 

\subsection{Supermassive Black Holes}
The high resolution imaging of Chandra was essential in studying SMBHs across a wide range of masses and cosmic times. Chandra explored the lowest mass SMBHs in nearby dwarf galaxies to the most massive ellipticals, and from the local universe to the cosmic dawn.

In the nearby universe, one of \chandra's key achievement was the direct detection of a binary SMBH system in an ongoing merger. Chandra observations of NGC 6240, a nearby merger remnant galaxy, detected not one but two luminous X-ray point sources associated with the optical core of the galaxies. Each of these nuclei showed characteristic signatures of accreting SMBHs, such as absorbed luminous X-ray emission and strong Fe K$\rm{\alpha}$ lines \cite{2003ApJ...582L..15K}. The two SMBHs, separated by $\approx$1~kpc, are orbiting each other and are predicted to coalesce in a few hundred million years. This discovery confirmed that galaxy mergers can bring SMBHs into close proximity, which can then result in their eventual merger. Chandra observations of other interacting galaxies detected many similar systems, clearly indicating that the formation of binary SMBHs is a common evolutionary stage of mergers \cite{2004ApJ...600..634B,2008MNRAS.386..105B}. By measuring the properties of binary SMBHs, Chandra can place constraints on these early phases of SMBH binary evolution, which at a later stage will be the source of low-frequency gravitational wave signals and will be targeted by nextâgeneration space-based observatories, such as LISA.

By the early 2000s, it was well established that virtually all Milky Way-sized and more massive galaxies host a SMBH. However, the presence of SMBHs in the much lower mass dwarf galaxies remained debated. That changed with the Chandra imaging of the dwarf starburst galaxy Henize\,2-10. By combining the Chandra Xâray data with highâresolution radio observations, a hard X-ray source was detected in the core of this galaxy, attributed to an accreting SMBH \cite{2011Natur.470...66R}. This detection showed that even low stellar mass galaxies can host and grow SMBHs. Following this discovery, Chandra studies of other dwarf galaxies identified further SMBH candidates, many of which were also identified in multi-wavelength observations in the radio, infrared, and optical bands \cite{2015ApJ...809L..14B,2017ApJ...836...20B}. 

Shortly after its launch, Chandra revolutionized our view of relativistic
jets. Its  first celestial target, the quasar PKS 0637-752, revealed a
$\gtrsim100$~kpc X-ray jet with four distinct knots aligned with previously
identified radio features \cite{2000ApJ...540L..69S}. Deep Chandra imaging of
the massive elliptical galaxy M\,87, then resolved its jet into multiple
synchrotron-emitting knots, which extending beyond $\sim5$~kpc
\cite{2002ApJ...564..683M}. Additionally, \chandra's High Resolution Camera
(HRC) could even track proper motions associated with the innermost knots
over a five year baseline, measuring relativistic speeds on sub-kpc scales
\cite{2019ApJ...879....8S}.  Many \chandra\ observations of quasar jets have
revealed extended emission coincident with radio structures, with the X-ray emission attributed to inverseâCompton (IC) scattering of CMB photons. Since the CMB energy density scales as \(\rho_{\rm CMB}\propto(1+z)^4\), this mechanism substantially boosts X-ray brightness at $z\gtrsim1$. Modeling the X-ray (IC) to radio (synchrotron) flux ratio provides both the magnetic field and its particle spectrum of the jet. From these parameters, the jetâs enthalpy flux can be computed (and can be as high as $\sim10^{46}\ \rm{erg \ s^{-1}}$), which directly links X-ray jet observations to powerful AGN feedback  \cite{2001MNRAS.321L...1C}. Over the past 25 years, \chandra\ has turned jet detections into precision experiments, using X-ray-bright outflows as laboratories for particle acceleration and as tracers of cosmic feedback.

Understanding the origin of the first SMBHs is a fundamental challenge.
Theoretical studies proposed two broad seeding pathways: ``light seeds'' from
the remnants of Population III stars ($M_{\rm seed}\sim10-100 \
\rm{M_{\odot}}$) \cite{2001ApJ...551L..27M} and ``heavy seeds'' from the
collapse of  pristine gas clouds 
($M_{\rm seed}\sim10^4-10^5 \ \rm{M_{\odot}}$) 
\cite{2006MNRAS.371.1813L}. To distinguish between these scenarios, SMBHs must be observed close to their formation epoch ($z\gtrsim9$). However, the faint and likely obscured nature of these first SMBHs makes this a daunting challenge. A breakthrough came when JWST identified a population of galaxies at $z=9-12$, and ultra-deep \chandra\ observations of this same field detected Xâray point sources associated with two of these galaxies: UHZ1 and GHZ9 at $z\approx10.1$ ($\approx460$ million years after the Big Bang) \cite{2024NatAs...8..126B,2023ApJ...955L..24G,2024ApJ...965L..21K} (Figure~\ref{fig:clusterfig1} bottom). The Xâray luminosities of the sources imply black hole masses of a few times $10^7 \ \rm{M_{\odot}}$, which corresponds to $\sim20â40\%$ of the stellar mass of their host galaxies, far exceeding the local ratio of $\sim0.2\%$. 
The existence of such massive black holes so early, combined with their
unusually high $M_{\rm BH}/M_{\rm \star}$ ratios (where $M_{\rm \star}$ is
the stellar mass), strongly favors the ``heavy seed'' scenario. These 
detections provide the most compelling evidence to date that some of the 
first SMBHs formed through the collapse of massive gas clouds.

\section{An Exploration of the Solar System and Stellar Habitability}

\noindent
From the earliest recognition of other planets within the Solar System, the notion that they might harbor some form of life has captivated the thoughts of philosophers and scientists alike. While our understanding of this question has become highly refined over long periods of investigation, efforts persist to identify signatures of life beyond Earth. Detailed studies of our own Solar System form the bedrock for understanding the conditions under which other forms of life, if they exist, would need to 
develop, and X-ray observations play a unique role in probing the physical conditions within the Solar System as well as the impact of solar radiation. These, in turn, inform studies of other stellar systems that host planets.

\subsection{Solar System Objects}
\noindent
The solar wind -- a stream of energetic electrons, ions, and entrained magnetic field -- provides the energy source for multiple sources of X-ray emission within the solar system. The early work that launched the field was directed at detecting X-ray from the moon, produced by fluorescence from particles impacting the surface \cite{1962PhRvL...9..439G}.

\begin{figure}[h]
\centering
\includegraphics[width=1.0\textwidth]{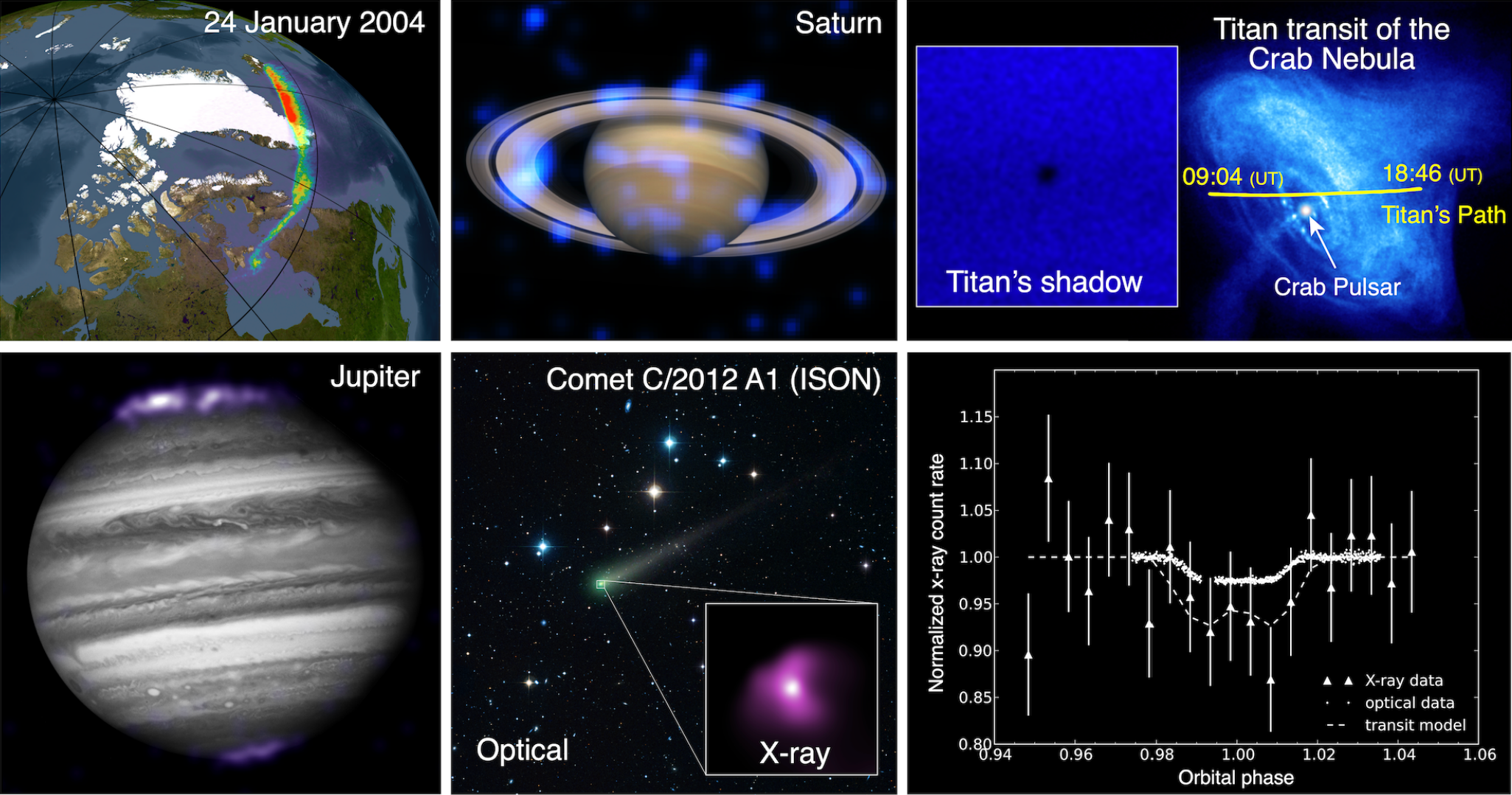}

\caption{
{\bf \chandra\ investigations of planets within the Solar System
and beyond.}
Counterclockwise from upper right: Saturn's moon Titan transiting
the Crab Nebula; Saturn, showing X-rays (blue) from disk and rings;
X-ray aurorae from Earth; aurorae from Jupiter; SWCX emission from
Comet C/2012 A1; X-ray transit of HD 189733. Error bars are $1\sigma$.
Bottom right plot adapted with permission from \cite{2013ApJ...773...62P},
AAAS. Credit: top left (X-ray aurorae), NASA/MSFC/CXC and
\cite{2007JASTP..69..179B}; top left (Earth model), NASA/GSFC/L.
Perkins and G. Shirah; top middle (X-ray), NASA/MSFC/CXC and
\cite{2005ApJ...627L..73B}; top middle (optical), NASA/ESA/STScI/AURA;
top right, NASA/CXC/Penn State and \cite{2004ApJ...607.1065M};
bottom left (X-ray), NASA/CXC/ SwRI/R. Gladstone; bottom left
(optical), NASA/ESA/Hubble Heritage (AURA/ STScI); bottom middle
(X-ray), NASA/CXC/Univ. CT and \cite{2016ApJ...818..199S}; bottom
middle (optical), DSS, Damian Peach
(\href{http://damianpeach.com}{damianpeach.com}).
}
\label{planetfig}
\end{figure}

To date, \chandra\ has detected X-rays From Venus, Earth, Mars,
Jupiter and its moons Io and Europa, Saturn and its rings, Pluto,
multiple comets, and even a transit of the Crab Nebula by Saturn's
moon Titan (see review by \cite{2019cxro.book....4D}).  The X-ray
emission results from a variety of processes -- charge exchange,
scattering of solar X-rays, fluorescence following solar X-ray
photoionization, and bremsstrahlung and line emission from collisions
with energetic particles. Which of these processes dominates is
dependent upon the conditions of both the solar wind and the
environment at the interaction site.

Figure \ref{planetfig} displays results from several of the Solar
System objects, with images of X-ray aurorae from the Earth (upper
left) and Jupiter (lower  left). Fluorescence and scattering of
solar X-rays are observed from both Saturn's disk and its rings
(middle top) while solar wind charge exchange (SWCX) dominates the
emission from Comet C/2012 (lower middle). The transit of the Crab
Nebula by Titan is shown at upper right; the shadow is larger than
the diameter of its solid surface, consistent with absorption by
Titan's upper atmosphere.

\subsection{Exoplanets}
\noindent
Despite contributing substantial advances in studies of X-rays from 
Solar System objects, the nearest extrasolar planets are far too distant to detect X-ray emission with \chandra. Yet the question of habitability of such exoplanets is heavily dependent upon the behavior of their hosts stars -- in particular as to whether or not planetary atmospheres can survive long episodes of X-ray radiation and flaring in young stellar systems, and whether the effects of planets around their host stars impact that stellar behavior. Studies with \chandra\ have yielded crucial information in this area.

Photoevaporation of exoplanet atmospheres, energized by extreme ultraviolet
and X-ray radiation from both quiescent and flaring stellar activity, is
an important consideration for habitability. While population studies
of stellar activity provide broad insight into the evolutionary impacts for
different star classes, an important question is whether the presence of
exoplanets themselves might impact the stellar activity, possibly through
tidal interactions. \chandra\ studies of HD 189733 AB and CoRoT-2 AM have
provided evidence for influence of hot Jupiters on the rotation activity of
the host stars, indicated by the planet-hosting stars in each of these
systems having higher X-ray fluxes than the similar coeval companion stars
\cite{2014A&A...565L...1P}. In addition, observations of the Hot Jupiter HD
189733b show a deeper transit than  observed in the optical
(Figure \ref{planetfig}, lower right), indicating the presence of a thin
outer atmosphere that is opaque to X-rays \cite{2013ApJ...773...62P}.
\chandra\ observations of the brown dwarf - M dwarf pair NLTT 41135/41166 --
two stars separated by only $2.5^{\prime\prime}$, and thus resolvable in X-rays only by \chandra\ -- show that N41135 is more than ten times as bright as N41136, a similar star but without an identified companion, suggesting that the presence of the planetary companion contributes to enhanced stellar activity \cite{2023MNRAS.524.5954I}.

More direct investigations of habitability center on quantitative assessments of how stellar activity impacts putative atmospheres.  \chandra\ and HST observations of Barnard's star, for example, show that the quiescent X-ray/UV flux is not sufficient to destroy the atmosphere of a habitable-zone terrestrial-like planet, but that sustained flaring at observed rates could result in rapid atmospheric mass loss for planets in this system or around other old M dwarfs \cite{2020AJ....160..237F}. Similarly, a study of the triple system LTT 1445, resolvable in X-rays only by \chandra, shows that planets orbiting LTT 1445A receive X-ray flux from all three stars in the system, but that even though C dominates the overall X-ray flux from the system, it does not impact the planets around A more than the emission from A itself \cite{2024A&A...687A.237R}. The conclusion is that, if starting with an Earth-equivalent amount of water, exoplanet LTT 1445Ad could maintain an atmosphere with water for over 1 Gyr.

Studies of larger stellar populations provide additional important
constraints on the question of habitability. X-ray studies of a
volume complete sample of 441 M/K/G stars using \chandra, XMM-Newton,
eROSITA, and ROSAT data \cite{2025A&A...694A..93Z} show that while
M-type stars show higher levels of X-ray activity than solar-type
stars, more than 60\% of all nearby M stars (and 80\% of early M
stars) have $L_x/L_{\rm bol}$ values within the normal range for G
stars (where $L_{\rm bol}$ is the bolometric luminosity), indicating
that such stars are potential hosts for habitable planets. On the
other hand, \chandra\ observations of 24,000 stars in 40 star-forming
regions reveal frequent and powerful flares from over 1000 stars
that could result in complete evaporation of any associated exoplanet
atmospheres over a time period of less than 5 million years
\cite{2021ApJ...916...32G}.

With an eye toward future telescopes that will provide capabilities for direct-imaging of exoplanets, \cite{2024ApJS..275....1B} investigated a sample of stars for which the habitable zone extends beyond the angular scale that will be resolvable by future telescope facilities, and for which Earth-sized planets with albedos similar to Earth would be detectable. From a total sample of 229 stars, X-ray data identify 29 systems with $(L_X/L_{\rm bol}) \lsim (L_X/L_{\rm bol})_\odot$, making these prime candidates for habitable exoplanet searches. However, nearly 70\% of the full sample currently lacks X-ray observations. \chandra\ studies of this sample would provide an extremely important pathfinder for future 
Habitable Worlds Observatory searches for habitable worlds.


\section{\chandra\ for the Future}
As of this writing, \chandra's high angular resolution capabilities remain unique.  No other X-ray observatory, operating or in development, has \chandra's ability to image cosmic sources on sub-arcsecond scales, or to cover $\sim 9$ orders of magnitude in flux.  These unparalleled features ensure that \chandra\ will be an integral element of observational astrophysics well into the future and offer a powerful complement to other observatories like HST, JWST, and the upcoming Nancy Grace Roman Space Telescope. The results from the observatory and its robust community have demonstrated the crucial role that \chandra\ observations play in addressing a vast array of problems in astrophysics -- including, as has been so often the case, many that have not yet been anticipated.

Examples of \chandra\ studies in the immediate future include an
ongoing \chandra\ Legacy Program (CLP) to complete observations of
all 74 galaxies from the multiwavelength Physics at High Angular
resolution in Nearby GalaxieS (PHANGS) survey to accompany
high-resolution data from AstroSAT, HST, Very Large Telescope(VLT)/Multi
Unit Spectroscopic Explorer (MUSE), Atacama Large Millimeter/submillimeter
Array (ALMA), and JWST that will probe the physics of baryon cycles
and feedback in nearby galaxies.  Deep observations of the Perseus
Cluster form the core of another CLP aimed at investigating the
transport and dissipation of energy within its core, providing
crucial information on AGN feedback from massive central galaxies
in cool core clusters.

Approved programs for future studies include major programs investigating exoplanets, impacts on habitability by stellar activity, and observations of Jupiter to study aurorae above the poles and to correlate X-ray behavior  with in-situ measurements from the Juno probe. Observations of 38 JWST-observed protostars in OMC-2 will trace the evolution of magnetic activity in the population.

\chandra\ will support an extended monitoring campaign of M87 with
the Event Horizon Telescope in a ``movie'' campaign to connect
variability in the near-horizon plasma to the behavior of the
larger-scale jet.  Among dozens of time domain studies, \chandra\
Target of Opportunity (TOO) programs will monitor and characterize
the X-ray counterparts of binary NS mergers and NS-BH mergers found
by Laser Interferometer Gravitational-Wave Observatory (LIGO)/Virgo
-- events for which we now know the rates are quite low, making it
crucial to maintain \chandra's capabilities to pursue such important
but rare opportunities. An additional TOO program will study the
anticipated thermonuclear event from the naked-eye recurrent nova
T Corona Borealis.  With crucial capabilities that will continued
to grow with the emergence of observations from the Vera Rubin
Observatory and the launch of Roman, \chandra\ will continue to
contribute broadly to time domain science.

The long time baseline provided by \chandra, coupled with the high angular resolution, offers important capabilities for dynamical studies. Future observations will include continued monitoring of SN 1987A as it transitions from a supernova into an SNR, measurements of structure changes in Cas A over the past 7\% of its lifetime, and proper motions of neutron stars to constrain kick velocities that probe asymmetries in SN explosions.

The longevity and success of the \chandra\ mission are due to the robust design of the observatory and to the diligence and dedication of its operations and science staff. The current health of the observatory is excellent, with no known issues that will prevent efficient and outstanding performance through the current decade, and likely well beyond. With a prognosis for continued outstanding scientific productivity, and a growing community of resourceful researchers, \chandra\ is well-positioned to continue its remarkable success as one of NASA's flagship observatories for exploring the universe for many years to come.

\backmatter





\begin{thebibliography}{100}
\expandafter\ifx\csname url\endcsname\relax
  \def\url#1{\burl{#1}}\fi
\expandafter\ifx\csname urlprefix\endcsname\relax\def\urlprefix{URL }\fi
\providecommand{\bibinfo}[2]{#2}
\providecommand{\eprint}[2][]{\url{#2}}
\providecommand{\doi}[1]{\url{https://doi.org/#1}}
\bibcommenthead

\bibitem{2019cxro.book.....W}
\bibinfo{author}{{Wilkes}, B.} \& \bibinfo{author}{{Tucker}, W.}
\newblock \emph{\bibinfo{title}{{The Chandra X-ray Observatory; Exploring the
  high energy universe}}}  (\bibinfo{publisher}{IOP Publishing, Bristol, UK},
  \bibinfo{year}{2019}).

\bibitem{2003SSRv..108..577F}
\bibinfo{author}{{Favata}, F.} \& \bibinfo{author}{{Micela}, G.}
\newblock \bibinfo{title}{{Stellar Coronal Astronomy}}.
\newblock \emph{\bibinfo{journal}{\ssr}} \textbf{\bibinfo{volume}{108}},
  \bibinfo{pages}{577--708} (\bibinfo{year}{2003}).

\bibitem{2014ApJ...787..109G}
\bibinfo{author}{{Getman}, K.~V.}, \bibinfo{author}{{Feigelson}, E.~D.} \&
  \bibinfo{author}{{Kuhn}, M.~A.}
\newblock \bibinfo{title}{{Core-Halo Age Gradients and Star Formation in the
  Orion Nebula and NGC 2024 Young Stellar Clusters}}.
\newblock \emph{\bibinfo{journal}{\apj}} \textbf{\bibinfo{volume}{787}},
  \bibinfo{pages}{109} (\bibinfo{year}{2014}).

\bibitem{2014ApJ...787..107K}
\bibinfo{author}{{Kuhn}, M.~A.} \emph{et~al.}
\newblock \bibinfo{title}{{The Spatial Structure of Young Stellar Clusters. I.
  Subclusters}}.
\newblock \emph{\bibinfo{journal}{\apj}} \textbf{\bibinfo{volume}{787}},
  \bibinfo{pages}{107} (\bibinfo{year}{2014}).

\bibitem{2015ApJ...802...60K}
\bibinfo{author}{{Kuhn}, M.~A.}, \bibinfo{author}{{Getman}, K.~V.} \&
  \bibinfo{author}{{Feigelson}, E.~D.}
\newblock \bibinfo{title}{{The Spatial Structure of Young Stellar Clusters. II.
  Total Young Stellar Populations}}.
\newblock \emph{\bibinfo{journal}{\apj}} \textbf{\bibinfo{volume}{802}},
  \bibinfo{pages}{60} (\bibinfo{year}{2015}).

\bibitem{2010ApJ...710.1835B}
\bibinfo{author}{{Brickhouse}, N.~S.}, \bibinfo{author}{{Cranmer}, S.~R.},
  \bibinfo{author}{{Dupree}, A.~K.}, \bibinfo{author}{{Luna}, G.~J.~M.} \&
  \bibinfo{author}{{Wolk}, S.}
\newblock \bibinfo{title}{{A Deep Chandra X-Ray Spectrum of the Accreting Young
  Star TW Hydrae}}.
\newblock \emph{\bibinfo{journal}{\apj}} \textbf{\bibinfo{volume}{710}},
  \bibinfo{pages}{1835--1847} (\bibinfo{year}{2010}).

\bibitem{2014prpl.conf..387A}
\bibinfo{author}{{Audard}, M.} \emph{et~al.}
\newblock \bibinfo{title}{ in \textit{Protostars and planets vi}} (eds
  \bibinfo{editor}{{Beuther}, H.}, \bibinfo{editor}{{Klessen}, R.~S.},
  \bibinfo{editor}{{Dullemond}, C.~P.} \& \bibinfo{editor}{{Henning}, T.})
  \emph{\bibinfo{booktitle}{{Episodic Accretion in Young Stars}}}
  (\bibinfo{publisher}{University of Arizona Press, Tucson, AZ},
  \bibinfo{year}{2014}).

\bibitem{2014A&A...570L..11L}
\bibinfo{author}{{Liebhart}, A.}, \bibinfo{author}{{G{\"u}del}, M.},
  \bibinfo{author}{{Skinner}, S.~L.} \& \bibinfo{author}{{Green}, J.}
\newblock \bibinfo{title}{{X-ray emission from an FU Orionis star in early
  outburst: HBC 722}}.
\newblock \emph{\bibinfo{journal}{\aap}} \textbf{\bibinfo{volume}{570}},
  \bibinfo{pages}{L11} (\bibinfo{year}{2014}).

\bibitem{2020AJ....159..221S}
\bibinfo{author}{{Skinner}, S.~L.} \& \bibinfo{author}{{G{\"u}del}, M.}
\newblock \bibinfo{title}{{Chandra Resolves the Double FU Orionis System RNO
  1B/1C in X-Rays}}.
\newblock \emph{\bibinfo{journal}{\aj}} \textbf{\bibinfo{volume}{159}},
  \bibinfo{pages}{221} (\bibinfo{year}{2020}).

\bibitem{2010ApJ...722.1654S}
\bibinfo{author}{{Skinner}, S.~L.}, \bibinfo{author}{{G{\"u}del}, M.},
  \bibinfo{author}{{Briggs}, K.~R.} \& \bibinfo{author}{{Lamzin}, S.~A.}
\newblock \bibinfo{title}{{Chandra Reveals Variable Multi-component X-ray
  Emission From FU Orionis}}.
\newblock \emph{\bibinfo{journal}{\apj}} \textbf{\bibinfo{volume}{722}},
  \bibinfo{pages}{1654--1665} (\bibinfo{year}{2010}).

\bibitem{2001Natur.413..708P}
\bibinfo{author}{{Pravdo}, S.~H.} \emph{et~al.}
\newblock \bibinfo{title}{{Discovery of X-rays from the protostellar outflow
  object HH2}}.
\newblock \emph{\bibinfo{journal}{\nat}} \textbf{\bibinfo{volume}{413}},
  \bibinfo{pages}{708--711} (\bibinfo{year}{2001}).

\bibitem{2023ApJS..269...13G}
\bibinfo{author}{{Guarcello}, M.~G.} \emph{et~al.}
\newblock \bibinfo{title}{{Photoevaporation and Close Encounters: How the
  Environment around Cygnus OB2 Affects the Evolution of Protoplanetary
  Disks}}.
\newblock \emph{\bibinfo{journal}{\apjs}} \textbf{\bibinfo{volume}{269}},
  \bibinfo{pages}{13} (\bibinfo{year}{2023}).

\bibitem{2010AsBio..10..751S}
\bibinfo{author}{{Segura}, A.}, \bibinfo{author}{{Walkowicz}, L.~M.},
  \bibinfo{author}{{Meadows}, V.}, \bibinfo{author}{{Kasting}, J.} \&
  \bibinfo{author}{{Hawley}, S.}
\newblock \bibinfo{title}{{The Effect of a Strong Stellar Flare on the
  Atmospheric Chemistry of an Earth-like Planet Orbiting an M Dwarf}}.
\newblock \emph{\bibinfo{journal}{Astrobiology}} \textbf{\bibinfo{volume}{10}},
  \bibinfo{pages}{751--771} (\bibinfo{year}{2010}).

\bibitem{2019NatAs...3..742A}
\bibinfo{author}{{Argiroffi}, C.} \emph{et~al.}
\newblock \bibinfo{title}{{A stellar flare-coronal mass ejection event revealed
  by X-ray plasma motions}}.
\newblock \emph{\bibinfo{journal}{\nastro}} \textbf{\bibinfo{volume}{3}},
  \bibinfo{pages}{742--748} (\bibinfo{year}{2019}).

\bibitem{2002ApJ...572..932P}
\bibinfo{author}{{Pooley}, D.} \emph{et~al.}
\newblock \bibinfo{title}{{X-Ray, Optical, and Radio Observations of the Type
  II Supernovae 1999em and 1998S}}.
\newblock \emph{\bibinfo{journal}{\apj}} \textbf{\bibinfo{volume}{572}},
  \bibinfo{pages}{932--943} (\bibinfo{year}{2002}).

\bibitem{2006ApJ...651.1005S}
\bibinfo{author}{{Soderberg}, A.~M.}, \bibinfo{author}{{Chevalier}, R.~A.},
  \bibinfo{author}{{Kulkarni}, S.~R.} \& \bibinfo{author}{{Frail}, D.~A.}
\newblock \bibinfo{title}{{The Radio and X-Ray Luminous SN 2003bg and the
  Circumstellar Density Variations around Radio Supernovae}}.
\newblock \emph{\bibinfo{journal}{\apj}} \textbf{\bibinfo{volume}{651}},
  \bibinfo{pages}{1005--1018} (\bibinfo{year}{2006}).

\bibitem{2007MNRAS.381..280M}
\bibinfo{author}{{Misra}, K.} \emph{et~al.}
\newblock \bibinfo{title}{{Type IIP supernova SN 2004et: a multiwavelength
  study in X-ray, optical and radio}}.
\newblock \emph{\bibinfo{journal}{\mnras}} \textbf{\bibinfo{volume}{381}},
  \bibinfo{pages}{280--292} (\bibinfo{year}{2007}).

\bibitem{2022ApJ...939..105B}
\bibinfo{author}{{Brethauer}, D.} \emph{et~al.}
\newblock \bibinfo{title}{{Seven Years of Coordinated Chandra-NuSTAR
  Observations of SN 2014C Unfold the Extreme Mass-loss History of Its Stellar
  Progenitor}}.
\newblock \emph{\bibinfo{journal}{\apj}} \textbf{\bibinfo{volume}{939}},
  \bibinfo{pages}{105} (\bibinfo{year}{2022}).

\bibitem{2022ApJ...930...57T}
\bibinfo{author}{{Thomas}, B.~P.} \emph{et~al.}
\newblock \bibinfo{title}{{Seven Years of SN 2014C: A Multiwavelength Synthesis
  of an Extraordinary Supernova}}.
\newblock \emph{\bibinfo{journal}{\apj}} \textbf{\bibinfo{volume}{930}},
  \bibinfo{pages}{57} (\bibinfo{year}{2022}).

\bibitem{2018MNRAS.478.5050M}
\bibinfo{author}{{Mauerhan}, J.~C.} \emph{et~al.}
\newblock \bibinfo{title}{{Stripped-envelope supernova SN 2004dk is now
  interacting with hydrogen-rich circumstellar material}}.
\newblock \emph{\bibinfo{journal}{\mnras}} \textbf{\bibinfo{volume}{478}},
  \bibinfo{pages}{5050--5055} (\bibinfo{year}{2018}).

\bibitem{2019ApJ...883..120P}
\bibinfo{author}{{Pooley}, D.} \emph{et~al.}
\newblock \bibinfo{title}{{Interaction of SN Ib 2004dk with a Previously
  Expelled Envelope}}.
\newblock \emph{\bibinfo{journal}{\apj}} \textbf{\bibinfo{volume}{883}},
  \bibinfo{pages}{120} (\bibinfo{year}{2019}).

\bibitem{2008ApJ...680.1359M}
\bibinfo{author}{{Miller}, J.~M.} \emph{et~al.}
\newblock \bibinfo{title}{{The Accretion Disk Wind in the Black Hole GRO
  J1655-40}}.
\newblock \emph{\bibinfo{journal}{\apj}} \textbf{\bibinfo{volume}{680}},
  \bibinfo{pages}{1359--1377} (\bibinfo{year}{2008}).

\bibitem{2006AdSpR..38.2937F}
\bibinfo{author}{{Fabbiano}, G.}
\newblock \bibinfo{title}{{X-ray populations in galaxies}}.
\newblock \emph{\bibinfo{journal}{\asr}} \textbf{\bibinfo{volume}{38}},
  \bibinfo{pages}{2937--2941} (\bibinfo{year}{2006}).

\bibitem{2012MNRAS.419.2095M}
\bibinfo{author}{{Mineo}, S.}, \bibinfo{author}{{Gilfanov}, M.} \&
  \bibinfo{author}{{Sunyaev}, R.}
\newblock \bibinfo{title}{{X-ray emission from star-forming galaxies - I.
  High-mass X-ray binaries}}.
\newblock \emph{\bibinfo{journal}{\mnras}} \textbf{\bibinfo{volume}{419}},
  \bibinfo{pages}{2095--2115} (\bibinfo{year}{2012}).

\bibitem{2011ApJ...729...12B}
\bibinfo{author}{{Boroson}, B.}, \bibinfo{author}{{Kim}, D.-W.} \&
  \bibinfo{author}{{Fabbiano}, G.}
\newblock \bibinfo{title}{{Revisiting with Chandra the Scaling Relations of the
  X-ray Emission Components (Binaries, Nuclei, and Hot Gas) of Early-type
  Galaxies}}.
\newblock \emph{\bibinfo{journal}{\apj}} \textbf{\bibinfo{volume}{729}},
  \bibinfo{pages}{12} (\bibinfo{year}{2011}).

\bibitem{2016ApJ...825....7L}
\bibinfo{author}{{Lehmer}, B.~D.} \emph{et~al.}
\newblock \bibinfo{title}{{The Evolution of Normal Galaxy X-Ray Emission
  through Cosmic History: Constraints from the 6 MS Chandra Deep Field-South}}.
\newblock \emph{\bibinfo{journal}{\apj}} \textbf{\bibinfo{volume}{825}},
  \bibinfo{pages}{7} (\bibinfo{year}{2016}).

\bibitem{2017ApJ...840...39M}
\bibinfo{author}{{Madau}, P.} \& \bibinfo{author}{{Fragos}, T.}
\newblock \bibinfo{title}{{Radiation Backgrounds at Cosmic Dawn: X-Rays from
  Compact Binaries}}.
\newblock \emph{\bibinfo{journal}{\apj}} \textbf{\bibinfo{volume}{840}},
  \bibinfo{pages}{39} (\bibinfo{year}{2017}).

\bibitem{2014MNRAS.443..678P}
\bibinfo{author}{{Pacucci}, F.}, \bibinfo{author}{{Mesinger}, A.},
  \bibinfo{author}{{Mineo}, S.} \& \bibinfo{author}{{Ferrara}, A.}
\newblock \bibinfo{title}{{The X-ray spectra of the first galaxies: 21 cm
  signatures}}.
\newblock \emph{\bibinfo{journal}{\mnras}} \textbf{\bibinfo{volume}{443}},
  \bibinfo{pages}{678--686} (\bibinfo{year}{2014}).

\bibitem{2013A&ARv..21...59I}
\bibinfo{author}{{Ivanova}, N.} \emph{et~al.}
\newblock \bibinfo{title}{{Common envelope evolution: where we stand and how we
  can move forward}}.
\newblock \emph{\bibinfo{journal}{\aapr}} \textbf{\bibinfo{volume}{21}},
  \bibinfo{pages}{59} (\bibinfo{year}{2013}).

\bibitem{1975ApJ...199L.143C}
\bibinfo{author}{{Clark}, G.~W.}
\newblock \bibinfo{title}{{X-ray binaries in globular clusters.}}
\newblock \emph{\bibinfo{journal}{\apjl}} \textbf{\bibinfo{volume}{199}},
  \bibinfo{pages}{L143--L145} (\bibinfo{year}{1975}).

\bibitem{1975Natur.253..698K}
\bibinfo{author}{{Katz}, J.~I.}
\newblock \bibinfo{title}{{Two kinds of stellar collapse}}.
\newblock \emph{\bibinfo{journal}{\nat}} \textbf{\bibinfo{volume}{253}},
  \bibinfo{pages}{698--699} (\bibinfo{year}{1975}).

\bibitem{1983ApJ...267L..83H}
\bibinfo{author}{{Hertz}, P.} \& \bibinfo{author}{{Grindlay}, J.~E.}
\newblock \bibinfo{title}{{X-ray evidence for white dwarf binaries in globular
  clusters.}}
\newblock \emph{\bibinfo{journal}{\apjl}} \textbf{\bibinfo{volume}{267}},
  \bibinfo{pages}{L83--L87} (\bibinfo{year}{1983}).

\bibitem{1983ApJ...275..105H}
\bibinfo{author}{{Hertz}, P.} \& \bibinfo{author}{{Grindlay}, J.~E.}
\newblock \bibinfo{title}{{An X-ray survey of globular clusters and their X-ray
  luminosity function.}}
\newblock \emph{\bibinfo{journal}{\apj}} \textbf{\bibinfo{volume}{275}},
  \bibinfo{pages}{105--119} (\bibinfo{year}{1983}).

\bibitem{2001A&A...368..137V}
\bibinfo{author}{{Verbunt}, F.}
\newblock \bibinfo{title}{{A census with ROSAT of low-luminosity X-ray sources
  in globular clusters}}.
\newblock \emph{\bibinfo{journal}{\aap}} \textbf{\bibinfo{volume}{368}},
  \bibinfo{pages}{137--159} (\bibinfo{year}{2001}).

\bibitem{2001Sci...292.2290G}
\bibinfo{author}{{Grindlay}, J.~E.}, \bibinfo{author}{{Heinke}, C.},
  \bibinfo{author}{{Edmonds}, P.~D.} \& \bibinfo{author}{{Murray}, S.~S.}
\newblock \bibinfo{title}{{High-Resolution X-ray Imaging of a Globular Cluster
  Core: Compact Binaries in 47Tuc}}.
\newblock \emph{\bibinfo{journal}{Science}} \textbf{\bibinfo{volume}{292}},
  \bibinfo{pages}{2290--2295} (\bibinfo{year}{2001}).

\bibitem{2001ApJ...563L..53G}
\bibinfo{author}{{Grindlay}, J.~E.}, \bibinfo{author}{{Heinke}, C.~O.},
  \bibinfo{author}{{Edmonds}, P.~D.}, \bibinfo{author}{{Murray}, S.~S.} \&
  \bibinfo{author}{{Cool}, A.~M.}
\newblock \bibinfo{title}{{Chandra Exposes the Core-collapsed Globular Cluster
  NGC 6397}}.
\newblock \emph{\bibinfo{journal}{\apjl}} \textbf{\bibinfo{volume}{563}},
  \bibinfo{pages}{L53--L56} (\bibinfo{year}{2001}).

\bibitem{2002ApJ...569..405P}
\bibinfo{author}{{Pooley}, D.} \emph{et~al.}
\newblock \bibinfo{title}{{Optical Identification of Multiple Faint X-Ray
  Sources in the Globular Cluster NGC 6752: Evidence for Numerous Cataclysmic
  Variables}}.
\newblock \emph{\bibinfo{journal}{\apj}} \textbf{\bibinfo{volume}{569}},
  \bibinfo{pages}{405--417} (\bibinfo{year}{2002}).

\bibitem{2002ApJ...578..405R}
\bibinfo{author}{{Rutledge}, R.~E.}, \bibinfo{author}{{Bildsten}, L.},
  \bibinfo{author}{{Brown}, E.~F.}, \bibinfo{author}{{Pavlov}, G.~G.} \&
  \bibinfo{author}{{Zavlin}, V.~E.}
\newblock \bibinfo{title}{{A Possible Transient Neutron Star in Quiescence in
  the Globular Cluster NGC 5139}}.
\newblock \emph{\bibinfo{journal}{\apj}} \textbf{\bibinfo{volume}{578}},
  \bibinfo{pages}{405--412} (\bibinfo{year}{2002}).

\bibitem{2002ApJ...573..184P}
\bibinfo{author}{{Pooley}, D.} \emph{et~al.}
\newblock \bibinfo{title}{{Chandra Observation of the Globular Cluster NGC 6440
  and the Nature of Cluster X-Ray Luminosity Functions}}.
\newblock \emph{\bibinfo{journal}{\apj}} \textbf{\bibinfo{volume}{573}},
  \bibinfo{pages}{184--190} (\bibinfo{year}{2002}).

\bibitem{2003ApJ...591L.131P}
\bibinfo{author}{{Pooley}, D.} \emph{et~al.}
\newblock \bibinfo{title}{{Dynamical Formation of Close Binary Systems in
  Globular Clusters}}.
\newblock \emph{\bibinfo{journal}{\apjl}} \textbf{\bibinfo{volume}{591}},
  \bibinfo{pages}{L131--L134} (\bibinfo{year}{2003}).

\bibitem{2003ApJ...598..501H}
\bibinfo{author}{{Heinke}, C.~O.} \emph{et~al.}
\newblock \bibinfo{title}{{Analysis of the Quiescent Low-Mass X-Ray Binary
  Population in Galactic Globular Clusters}}.
\newblock \emph{\bibinfo{journal}{\apj}} \textbf{\bibinfo{volume}{598}},
  \bibinfo{pages}{501--515} (\bibinfo{year}{2003}).

\bibitem{2003A&A...400..521G}
\bibinfo{author}{{Gendre}, B.}, \bibinfo{author}{{Barret}, D.} \&
  \bibinfo{author}{{Webb}, N.~A.}
\newblock \bibinfo{title}{{An XMM-Newton observation of the globular cluster
  Omega Centauri}}.
\newblock \emph{\bibinfo{journal}{\aap}} \textbf{\bibinfo{volume}{400}},
  \bibinfo{pages}{521--531} (\bibinfo{year}{2003}).

\bibitem{2006ApJ...646L.143P}
\bibinfo{author}{{Pooley}, D.} \& \bibinfo{author}{{Hut}, P.}
\newblock \bibinfo{title}{{Dynamical Formation of Close Binaries in Globular
  Clusters: Cataclysmic Variables}}.
\newblock \emph{\bibinfo{journal}{\apjl}} \textbf{\bibinfo{volume}{646}},
  \bibinfo{pages}{L143--L146} (\bibinfo{year}{2006}).

\bibitem{2017ApJ...848L..25H}
\bibinfo{author}{{Haggard}, D.} \emph{et~al.}
\newblock \bibinfo{title}{{A Deep Chandra X-Ray Study of Neutron Star
  Coalescence GW170817}}.
\newblock \emph{\bibinfo{journal}{\apjl}} \textbf{\bibinfo{volume}{848}},
  \bibinfo{pages}{L25} (\bibinfo{year}{2017}).

\bibitem{2017ApJ...848L..20M}
\bibinfo{author}{{Margutti}, R.} \emph{et~al.}
\newblock \bibinfo{title}{{The Electromagnetic Counterpart of the Binary
  Neutron Star Merger LIGO/Virgo GW170817. V. Rising X-Ray Emission from an
  Off-axis Jet}}.
\newblock \emph{\bibinfo{journal}{\apjl}} \textbf{\bibinfo{volume}{848}},
  \bibinfo{pages}{L20} (\bibinfo{year}{2017}).

\bibitem{2017Natur.551...71T}
\bibinfo{author}{{Troja}, E.} \emph{et~al.}
\newblock \bibinfo{title}{{The X-ray counterpart to the gravitational-wave
  event GW170817}}.
\newblock \emph{\bibinfo{journal}{\nat}} \textbf{\bibinfo{volume}{551}},
  \bibinfo{pages}{71--74} (\bibinfo{year}{2017}).

\bibitem{2020MNRAS.498.5643T}
\bibinfo{author}{{Troja}, E.} \emph{et~al.}
\newblock \bibinfo{title}{{A thousand days after the merger: Continued X-ray
  emission from GW170817}}.
\newblock \emph{\bibinfo{journal}{\mnras}} \textbf{\bibinfo{volume}{498}},
  \bibinfo{pages}{5643--5651} (\bibinfo{year}{2020}).

\bibitem{2017PhRvL.119p1101A}
\bibinfo{author}{{Abbott}, B.~P.} \emph{et~al.}
\newblock \bibinfo{title}{{GW170817: Observation of Gravitational Waves from a
  Binary Neutron Star Inspiral}}.
\newblock \emph{\bibinfo{journal}{\prl}} \textbf{\bibinfo{volume}{119}},
  \bibinfo{pages}{161101} (\bibinfo{year}{2017}).

\bibitem{2019MNRAS.483.1912P}
\bibinfo{author}{{Piro}, L.} \emph{et~al.}
\newblock \bibinfo{title}{{A long-lived neutron star merger remnant in
  GW170817: constraints and clues from X-ray observations}}.
\newblock \emph{\bibinfo{journal}{\mnras}} \textbf{\bibinfo{volume}{483}},
  \bibinfo{pages}{1912--1921} (\bibinfo{year}{2019}).

\bibitem{2000ApJ...528L.109H}
\bibinfo{author}{{Hughes}, J.~P.}, \bibinfo{author}{{Rakowski}, C.~E.},
  \bibinfo{author}{{Burrows}, D.~N.} \& \bibinfo{author}{{Slane}, P.~O.}
\newblock \bibinfo{title}{{Nucleosynthesis and Mixing in Cassiopeia A}}.
\newblock \emph{\bibinfo{journal}{\apjl}} \textbf{\bibinfo{volume}{528}},
  \bibinfo{pages}{L109--L113} (\bibinfo{year}{2000}).

\bibitem{2012ApJ...746..130H}
\bibinfo{author}{{Hwang}, U.} \& \bibinfo{author}{{Laming}, J.~M.}
\newblock \bibinfo{title}{{A Chandra X-Ray Survey of Ejecta in the Cassiopeia A
  Supernova Remnant}}.
\newblock \emph{\bibinfo{journal}{\apj}} \textbf{\bibinfo{volume}{746}},
  \bibinfo{pages}{130} (\bibinfo{year}{2012}).

\bibitem{2024ApJ...965L..27M}
\bibinfo{author}{{Milisavljevic}, D.} \emph{et~al.}
\newblock \bibinfo{title}{{A JWST Survey of the Supernova Remnant Cassiopeia
  A}}.
\newblock \emph{\bibinfo{journal}{\apjl}} \textbf{\bibinfo{volume}{965}},
  \bibinfo{pages}{L27} (\bibinfo{year}{2024}).

\bibitem{2023MNRAS.518.2775S}
\bibinfo{author}{{Shternin}, P.~S.}, \bibinfo{author}{{Ofengeim}, D.~D.},
  \bibinfo{author}{{Heinke}, C.~O.} \& \bibinfo{author}{{Ho}, W. C.~G.}
\newblock \bibinfo{title}{{Constraints on neutron star superfluidity from the
  cooling neutron star in Cassiopeia A using all Chandra ACIS-S observations}}.
\newblock \emph{\bibinfo{journal}{\mnras}} \textbf{\bibinfo{volume}{518}},
  \bibinfo{pages}{2775--2793} (\bibinfo{year}{2023}).

\bibitem{2016ApJ...829...40F}
\bibinfo{author}{{Frank}, K.~A.} \emph{et~al.}
\newblock \bibinfo{title}{{Chandra Observes the End of an Era in SN 1987A}}.
\newblock \emph{\bibinfo{journal}{\apj}} \textbf{\bibinfo{volume}{829}},
  \bibinfo{pages}{40} (\bibinfo{year}{2016}).

\bibitem{2024ApJ...966..147R}
\bibinfo{author}{{Ravi}, A.~P.} \emph{et~al.}
\newblock \bibinfo{title}{{Latest Evolution of the X-Ray Remnant of SN 1987A:
  Beyond the Inner Ring}}.
\newblock \emph{\bibinfo{journal}{\apj}} \textbf{\bibinfo{volume}{966}},
  \bibinfo{pages}{147} (\bibinfo{year}{2024}).

\bibitem{2007ApJ...668L.135R}
\bibinfo{author}{{Reynolds}, S.~P.} \emph{et~al.}
\newblock \bibinfo{title}{{A Deep Chandra Observation of Kepler's Supernova
  Remnant: A Type Ia Event with Circumstellar Interaction}}.
\newblock \emph{\bibinfo{journal}{\apjl}} \textbf{\bibinfo{volume}{668}},
  \bibinfo{pages}{L135--L138} (\bibinfo{year}{2007}).

\bibitem{2011ApJ...732..114L}
\bibinfo{author}{{Lopez}, L.~A.}, \bibinfo{author}{{Ramirez-Ruiz}, E.},
  \bibinfo{author}{{Huppenkothen}, D.}, \bibinfo{author}{{Badenes}, C.} \&
  \bibinfo{author}{{Pooley}, D.~A.}
\newblock \bibinfo{title}{{Using the X-ray Morphology of Young Supernova
  Remnants to Constrain Explosion Type, Ejecta Distribution, and Chemical
  Mixing}}.
\newblock \emph{\bibinfo{journal}{\apj}} \textbf{\bibinfo{volume}{732}},
  \bibinfo{pages}{114} (\bibinfo{year}{2011}).

\bibitem{2014ApJ...785L..27Y}
\bibinfo{author}{{Yamaguchi}, H.} \emph{et~al.}
\newblock \bibinfo{title}{{Discriminating the Progenitor Type of Supernova
  Remnants with Iron K-shell Emission}}.
\newblock \emph{\bibinfo{journal}{\apjl}} \textbf{\bibinfo{volume}{785}},
  \bibinfo{pages}{L27} (\bibinfo{year}{2014}).

\bibitem{2011Ap&SS.336..257R}
\bibinfo{author}{{Reynolds}, S.~P.}
\newblock \bibinfo{title}{{Particle acceleration in supernova-remnant shocks}}.
\newblock \emph{\bibinfo{journal}{\apss}} \textbf{\bibinfo{volume}{336}},
  \bibinfo{pages}{257--262} (\bibinfo{year}{2011}).

\bibitem{2012A&ARv..20...49V}
\bibinfo{author}{{Vink}, J.}
\newblock \bibinfo{title}{{Supernova remnants: the X-ray perspective}}.
\newblock \emph{\bibinfo{journal}{\aapr}} \textbf{\bibinfo{volume}{20}},
  \bibinfo{pages}{49} (\bibinfo{year}{2012}).

\bibitem{2024Galax..12...59S}
\bibinfo{author}{{Slane}, P.}, \bibinfo{author}{{Ferrazzoli}, R.},
  \bibinfo{author}{{Zhou}, P.} \& \bibinfo{author}{{Vink}, J.}
\newblock \bibinfo{title}{{Probing Magnetic Fields in Young Supernova Remnants
  with IXPE}}.
\newblock \emph{\bibinfo{journal}{Galaxies}} \textbf{\bibinfo{volume}{12}},
  \bibinfo{pages}{59} (\bibinfo{year}{2024}).

\bibitem{2006ApJS..163..335F}
\bibinfo{author}{{Fryer}, C.~L.} \& \bibinfo{author}{{Kusenko}, A.}
\newblock \bibinfo{title}{{Effects of Neutrino-driven Kicks on the Supernova
  Explosion Mechanism}}.
\newblock \emph{\bibinfo{journal}{\apjs}} \textbf{\bibinfo{volume}{163}},
  \bibinfo{pages}{335--343} (\bibinfo{year}{2006}).

\bibitem{2001ApJ...549.1111L}
\bibinfo{author}{{Lai}, D.}, \bibinfo{author}{{Chernoff}, D.~F.} \&
  \bibinfo{author}{{Cordes}, J.~M.}
\newblock \bibinfo{title}{{Pulsar Jets: Implications for Neutron Star Kicks and
  Initial Spins}}.
\newblock \emph{\bibinfo{journal}{\apj}} \textbf{\bibinfo{volume}{549}},
  \bibinfo{pages}{1111--1118} (\bibinfo{year}{2001}).

\bibitem{1998Natur.393..139S}
\bibinfo{author}{{Spruit}, H.} \& \bibinfo{author}{{Phinney}, E.~S.}
\newblock \bibinfo{title}{{Birth kicks as the origin of pulsar rotation}}.
\newblock \emph{\bibinfo{journal}{\nat}} \textbf{\bibinfo{volume}{393}},
  \bibinfo{pages}{139--141} (\bibinfo{year}{1998}).

\bibitem{2006A&A...457..963S}
\bibinfo{author}{{Scheck}, L.}, \bibinfo{author}{{Kifonidis}, K.},
  \bibinfo{author}{{Janka}, H.~T.} \& \bibinfo{author}{{M{\"u}ller}, E.}
\newblock \bibinfo{title}{{Multidimensional supernova simulations with
  approximative neutrino transport. I. Neutron star kicks and the anisotropy of
  neutrino-driven explosions in two spatial dimensions}}.
\newblock \emph{\bibinfo{journal}{\aap}} \textbf{\bibinfo{volume}{457}},
  \bibinfo{pages}{963--986} (\bibinfo{year}{2006}).

\bibitem{2013A&A...552A.126W}
\bibinfo{author}{{Wongwathanarat}, A.}, \bibinfo{author}{{Janka}, H.~T.} \&
  \bibinfo{author}{{M{\"u}ller}, E.}
\newblock \bibinfo{title}{{Three-dimensional neutrino-driven supernovae:
  Neutron star kicks, spins, and asymmetric ejection of nucleosynthesis
  products}}.
\newblock \emph{\bibinfo{journal}{\aap}} \textbf{\bibinfo{volume}{552}},
  \bibinfo{pages}{A126} (\bibinfo{year}{2013}).

\bibitem{2017ApJ...844...84H}
\bibinfo{author}{{Holland-Ashford}, T.}, \bibinfo{author}{{Lopez}, L.~A.},
  \bibinfo{author}{{Auchettl}, K.}, \bibinfo{author}{{Temim}, T.} \&
  \bibinfo{author}{{Ramirez-Ruiz}, E.}
\newblock \bibinfo{title}{{Comparing Neutron Star Kicks to Supernova Remnant
  Asymmetries}}.
\newblock \emph{\bibinfo{journal}{\apj}} \textbf{\bibinfo{volume}{844}},
  \bibinfo{pages}{84} (\bibinfo{year}{2017}).

\bibitem{2000ApJ...536L..81W}
\bibinfo{author}{{Weisskopf}, M.~C.} \emph{et~al.}
\newblock \bibinfo{title}{{Discovery of Spatial and Spectral Structure in the
  X-Ray Emission from the Crab Nebula}}.
\newblock \emph{\bibinfo{journal}{\apjl}} \textbf{\bibinfo{volume}{536}},
  \bibinfo{pages}{L81--L84} (\bibinfo{year}{2000}).

\bibitem{2004ApJ...616..403S}
\bibinfo{author}{{Slane}, P.}, \bibinfo{author}{{Helfand}, D.~J.},
  \bibinfo{author}{{van der Swaluw}, E.} \& \bibinfo{author}{{Murray}, S.~S.}
\newblock \bibinfo{title}{{New Constraints on the Structure and Evolution of
  the Pulsar Wind Nebula 3C 58}}.
\newblock \emph{\bibinfo{journal}{\apj}} \textbf{\bibinfo{volume}{616}},
  \bibinfo{pages}{403--413} (\bibinfo{year}{2004}).

\bibitem{2004ApJ...601..479N}
\bibinfo{author}{{Ng}, C.~Y.} \& \bibinfo{author}{{Romani}, R.~W.}
\newblock \bibinfo{title}{{Fitting Pulsar Wind Tori}}.
\newblock \emph{\bibinfo{journal}{\apj}} \textbf{\bibinfo{volume}{601}},
  \bibinfo{pages}{479--484} (\bibinfo{year}{2004}).

\bibitem{2008AIPC..983..171K}
\bibinfo{author}{{Kargaltsev}, O.} \& \bibinfo{author}{{Pavlov}, G.~G.}
\newblock \bibinfo{title}{ in \textit{40 years of pulsars: Millisecond pulsars,
  magnetars and more}} (eds \bibinfo{editor}{{Bassa}, C.},
  \bibinfo{editor}{{Wang}, Z.}, \bibinfo{editor}{{Cumming}, A.} \&
  \bibinfo{editor}{{Kaspi}, V.~M.}) \emph{\bibinfo{booktitle}{{Pulsar Wind
  Nebulae in the Chandra Era}}}  (\bibinfo{publisher}{AIP, Melville, NY},
  \bibinfo{year}{2008}).

\bibitem{2007ApJ...670L.121P}
\bibinfo{author}{{Park}, S.} \emph{et~al.}
\newblock \bibinfo{title}{{A Half-Megasecond Chandra Observation of the
  Oxygen-rich Supernova Remnant G292.0+1.8}}.
\newblock \emph{\bibinfo{journal}{\apjl}} \textbf{\bibinfo{volume}{670}},
  \bibinfo{pages}{L121--L124} (\bibinfo{year}{2007}).

\bibitem{2019ApJ...878L..19T}
\bibinfo{author}{{Temim}, T.} \emph{et~al.}
\newblock \bibinfo{title}{{Probing the Innermost Ejecta Layers in Supernova
  Remnant Kes 75: Implications for the Supernova Progenitor}}.
\newblock \emph{\bibinfo{journal}{\apjl}} \textbf{\bibinfo{volume}{878}},
  \bibinfo{pages}{L19} (\bibinfo{year}{2019}).

\bibitem{2008A&A...490L...3B}
\bibinfo{author}{{Bandiera}, R.}
\newblock \bibinfo{title}{{On the X-ray feature associated with the Guitar
  nebula}}.
\newblock \emph{\bibinfo{journal}{\aap}} \textbf{\bibinfo{volume}{490}},
  \bibinfo{pages}{L3--L6} (\bibinfo{year}{2008}).

\bibitem{2017ApJS..228....2L}
\bibinfo{author}{{Luo}, B.} \emph{et~al.}
\newblock \bibinfo{title}{{The Chandra Deep Field-South Survey: 7 Ms Source
  Catalogs}}.
\newblock \emph{\bibinfo{journal}{\apjs}} \textbf{\bibinfo{volume}{228}},
  \bibinfo{pages}{2} (\bibinfo{year}{2017}).

\bibitem{2006ApJ...645...95H}
\bibinfo{author}{{Hickox}, R.~C.} \& \bibinfo{author}{{Markevitch}, M.}
\newblock \bibinfo{title}{{Absolute Measurement of the Unresolved Cosmic X-Ray
  Background in the 0.5-8 keV Band with Chandra}}.
\newblock \emph{\bibinfo{journal}{\apj}} \textbf{\bibinfo{volume}{645}},
  \bibinfo{pages}{95--114} (\bibinfo{year}{2006}).

\bibitem{2007A&A...463...79G}
\bibinfo{author}{{Gilli}, R.}, \bibinfo{author}{{Comastri}, A.} \&
  \bibinfo{author}{{Hasinger}, G.}
\newblock \bibinfo{title}{{The synthesis of the cosmic X-ray background in the
  Chandra and XMM-Newton era}}.
\newblock \emph{\bibinfo{journal}{\aap}} \textbf{\bibinfo{volume}{463}},
  \bibinfo{pages}{79--96} (\bibinfo{year}{2007}).

\bibitem{2008ApJ...679..118S}
\bibinfo{author}{{Silverman}, J.~D.} \emph{et~al.}
\newblock \bibinfo{title}{{The Luminosity Function of X-Ray-selected Active
  Galactic Nuclei: Evolution of Supermassive Black Holes at High Redshift}}.
\newblock \emph{\bibinfo{journal}{\apj}} \textbf{\bibinfo{volume}{679}},
  \bibinfo{pages}{118--139} (\bibinfo{year}{2008}).

\bibitem{2004NuPhS.132...86H}
\bibinfo{author}{{Hasinger}, G.}
\newblock \bibinfo{title}{{The X-ray background and AGNs}}.
\newblock \emph{\bibinfo{journal}{\npbps}} \textbf{\bibinfo{volume}{132}},
  \bibinfo{pages}{86--96} (\bibinfo{year}{2004}).

\bibitem{2015A&ARv..23....1B}
\bibinfo{author}{{Brandt}, W.~N.} \& \bibinfo{author}{{Alexander}, D.~M.}
\newblock \bibinfo{title}{{Cosmic X-ray surveys of distant active galaxies. The
  demographics, physics, and ecology of growing supermassive black holes}}.
\newblock \emph{\bibinfo{journal}{\aapr}} \textbf{\bibinfo{volume}{23}},
  \bibinfo{pages}{1} (\bibinfo{year}{2015}).

\bibitem{1986PASJ...38..121K}
\bibinfo{author}{{Koyama}, K.}, \bibinfo{author}{{Makishima}, K.},
  \bibinfo{author}{{Tanaka}, Y.} \& \bibinfo{author}{{Tsunemi}, H.}
\newblock \bibinfo{title}{{Thermal X-ray emission with intense 6.7-keV iron
  line from the galactic ridge.}}
\newblock \emph{\bibinfo{journal}{\pasj}} \textbf{\bibinfo{volume}{38}},
  \bibinfo{pages}{121--131} (\bibinfo{year}{1986}).

\bibitem{2007A&A...473..857R}
\bibinfo{author}{{Revnivtsev}, M.}, \bibinfo{author}{{Vikhlinin}, A.} \&
  \bibinfo{author}{{Sazonov}, S.}
\newblock \bibinfo{title}{{Resolving the Galactic ridge X-ray background}}.
\newblock \emph{\bibinfo{journal}{\aap}} \textbf{\bibinfo{volume}{473}},
  \bibinfo{pages}{857--862} (\bibinfo{year}{2007}).

\bibitem{2009Natur.458.1142R}
\bibinfo{author}{{Revnivtsev}, M.} \emph{et~al.}
\newblock \bibinfo{title}{{Discrete sources as the origin of the Galactic X-ray
  ridge emission}}.
\newblock \emph{\bibinfo{journal}{\nat}} \textbf{\bibinfo{volume}{458}},
  \bibinfo{pages}{1142--1144} (\bibinfo{year}{2009}).

\bibitem{2007A&A...473..783R}
\bibinfo{author}{{Revnivtsev}, M.}, \bibinfo{author}{{Churazov}, E.},
  \bibinfo{author}{{Sazonov}, S.}, \bibinfo{author}{{Forman}, W.} \&
  \bibinfo{author}{{Jones}, C.}
\newblock \bibinfo{title}{{X-ray emission from the stellar population in M
  32}}.
\newblock \emph{\bibinfo{journal}{\aap}} \textbf{\bibinfo{volume}{473}},
  \bibinfo{pages}{783--789} (\bibinfo{year}{2007}).

\bibitem{2008MNRAS.388...56B}
\bibinfo{author}{{Bogd{\'a}n}, {\'A}.} \& \bibinfo{author}{{Gilfanov}, M.}
\newblock \bibinfo{title}{{Unresolved emission and ionized gas in the bulge of
  M31}}.
\newblock \emph{\bibinfo{journal}{\mnras}} \textbf{\bibinfo{volume}{388}},
  \bibinfo{pages}{56--66} (\bibinfo{year}{2008}).

\bibitem{2009ApJ...697.2030S}
\bibinfo{author}{{Strickland}, D.~K.} \& \bibinfo{author}{{Heckman}, T.~M.}
\newblock \bibinfo{title}{{Supernova Feedback Efficiency and Mass Loading in
  the Starburst and Galactic Superwind Exemplar M82}}.
\newblock \emph{\bibinfo{journal}{\apj}} \textbf{\bibinfo{volume}{697}},
  \bibinfo{pages}{2030--2056} (\bibinfo{year}{2009}).

\bibitem{2015ApJ...803L..21V}
\bibinfo{author}{{Voit}, G.~M.} \emph{et~al.}
\newblock \bibinfo{title}{{Supernova Sweeping and Black Hole Feedback in
  Elliptical Galaxies}}.
\newblock \emph{\bibinfo{journal}{\apjl}} \textbf{\bibinfo{volume}{803}},
  \bibinfo{pages}{L21} (\bibinfo{year}{2015}).

\bibitem{2005ApJ...624..751Y}
\bibinfo{author}{{Yao}, Y.} \& \bibinfo{author}{{Wang}, Q.~D.}
\newblock \bibinfo{title}{{X-Ray Absorption Line Spectroscopy of the Galactic
  Hot Interstellar Medium}}.
\newblock \emph{\bibinfo{journal}{\apj}} \textbf{\bibinfo{volume}{624}},
  \bibinfo{pages}{751--764} (\bibinfo{year}{2005}).

\bibitem{2012ApJ...756L...8G}
\bibinfo{author}{{Gupta}, A.}, \bibinfo{author}{{Mathur}, S.},
  \bibinfo{author}{{Krongold}, Y.}, \bibinfo{author}{{Nicastro}, F.} \&
  \bibinfo{author}{{Galeazzi}, M.}
\newblock \bibinfo{title}{{A Huge Reservoir of Ionized Gas around the Milky
  Way: Accounting for the Missing Mass?}}
\newblock \emph{\bibinfo{journal}{\apjl}} \textbf{\bibinfo{volume}{756}},
  \bibinfo{pages}{L8} (\bibinfo{year}{2012}).

\bibitem{2013ApJ...772...97B}
\bibinfo{author}{{Bogd{\'a}n}, {\'A}.} \emph{et~al.}
\newblock \bibinfo{title}{{Hot X-Ray Coronae around Massive Spiral Galaxies: A
  Unique Probe of Structure Formation Models}}.
\newblock \emph{\bibinfo{journal}{\apj}} \textbf{\bibinfo{volume}{772}},
  \bibinfo{pages}{97} (\bibinfo{year}{2013}).

\bibitem{2011ApJ...737...22A}
\bibinfo{author}{{Anderson}, M.~E.} \& \bibinfo{author}{{Bregman}, J.~N.}
\newblock \bibinfo{title}{{Detection of a Hot Gaseous Halo around the Giant
  Spiral Galaxy NGC 1961}}.
\newblock \emph{\bibinfo{journal}{\apj}} \textbf{\bibinfo{volume}{737}},
  \bibinfo{pages}{22} (\bibinfo{year}{2011}).

\bibitem{1978MNRAS.183..341W}
\bibinfo{author}{{White}, S.~D.~M.} \& \bibinfo{author}{{Rees}, M.~J.}
\newblock \bibinfo{title}{{Core condensation in heavy halos: a two-stage theory
  for galaxy formation and clustering.}}
\newblock \emph{\bibinfo{journal}{\mnras}} \textbf{\bibinfo{volume}{183}},
  \bibinfo{pages}{341--358} (\bibinfo{year}{1978}).

\bibitem{1991ApJ...379...52W}
\bibinfo{author}{{White}, S. D.~M.} \& \bibinfo{author}{{Frenk}, C.~S.}
\newblock \bibinfo{title}{{Galaxy Formation through Hierarchical Clustering}}.
\newblock \emph{\bibinfo{journal}{\apj}} \textbf{\bibinfo{volume}{379}},
  \bibinfo{pages}{52} (\bibinfo{year}{1991}).

\bibitem{2004ApJ...606..819M}
\bibinfo{author}{{Markevitch}, M.} \emph{et~al.}
\newblock \bibinfo{title}{{Direct Constraints on the Dark Matter
  Self-Interaction Cross Section from the Merging Galaxy Cluster 1E 0657-56}}.
\newblock \emph{\bibinfo{journal}{\apj}} \textbf{\bibinfo{volume}{606}},
  \bibinfo{pages}{819--824} (\bibinfo{year}{2004}).

\bibitem{2006ApJ...648L.109C}
\bibinfo{author}{{Clowe}, D.} \emph{et~al.}
\newblock \bibinfo{title}{{A Direct Empirical Proof of the Existence of Dark
  Matter}}.
\newblock \emph{\bibinfo{journal}{\apjl}} \textbf{\bibinfo{volume}{648}},
  \bibinfo{pages}{L109--L113} (\bibinfo{year}{2006}).

\bibitem{2008ApJ...679.1173R}
\bibinfo{author}{{Randall}, S.~W.}, \bibinfo{author}{{Markevitch}, M.},
  \bibinfo{author}{{Clowe}, D.}, \bibinfo{author}{{Gonzalez}, A.~H.} \&
  \bibinfo{author}{{Brada{\v{c}}}, M.}
\newblock \bibinfo{title}{{Constraints on the Self-Interaction Cross Section of
  Dark Matter from Numerical Simulations of the Merging Galaxy Cluster 1E
  0657-56}}.
\newblock \emph{\bibinfo{journal}{\apj}} \textbf{\bibinfo{volume}{679}},
  \bibinfo{pages}{1173--1180} (\bibinfo{year}{2008}).

\bibitem{2004MNRAS.353..457A}
\bibinfo{author}{{Allen}, S.~W.}, \bibinfo{author}{{Schmidt}, R.~W.},
  \bibinfo{author}{{Ebeling}, H.}, \bibinfo{author}{{Fabian}, A.~C.} \&
  \bibinfo{author}{{van Speybroeck}, L.}
\newblock \bibinfo{title}{{Constraints on dark energy from Chandra observations
  of the largest relaxed galaxy clusters}}.
\newblock \emph{\bibinfo{journal}{\mnras}} \textbf{\bibinfo{volume}{353}},
  \bibinfo{pages}{457--467} (\bibinfo{year}{2004}).

\bibitem{2009ApJ...692.1060V}
\bibinfo{author}{{Vikhlinin}, A.} \emph{et~al.}
\newblock \bibinfo{title}{{Chandra Cluster Cosmology Project III: Cosmological
  Parameter Constraints}}.
\newblock \emph{\bibinfo{journal}{\apj}} \textbf{\bibinfo{volume}{692}},
  \bibinfo{pages}{1060--1074} (\bibinfo{year}{2009}).

\bibitem{2005ApJ...635..894F}
\bibinfo{author}{{Forman}, W.} \emph{et~al.}
\newblock \bibinfo{title}{{Reflections of Active Galactic Nucleus Outbursts in
  the Gaseous Atmosphere of M87}}.
\newblock \emph{\bibinfo{journal}{\apj}} \textbf{\bibinfo{volume}{635}},
  \bibinfo{pages}{894--906} (\bibinfo{year}{2005}).

\bibitem{2024MNRAS.535.2173F}
\bibinfo{author}{{Fabian}, A.~C.} \emph{et~al.}
\newblock \bibinfo{title}{{Hidden cooling flows - IV. More details on Centaurus
  and the efficiency of AGN feedback in clusters}}.
\newblock \emph{\bibinfo{journal}{\mnras}} \textbf{\bibinfo{volume}{535}},
  \bibinfo{pages}{2173--2188} (\bibinfo{year}{2024}).

\bibitem{2000MNRAS.318L..65F}
\bibinfo{author}{{Fabian}, A.~C.} \emph{et~al.}
\newblock \bibinfo{title}{{Chandra imaging of the complex X-ray core of the
  Perseus cluster}}.
\newblock \emph{\bibinfo{journal}{\mnras}} \textbf{\bibinfo{volume}{318}},
  \bibinfo{pages}{L65--L68} (\bibinfo{year}{2000}).

\bibitem{2003MNRAS.344L..43F}
\bibinfo{author}{{Fabian}, A.~C.} \emph{et~al.}
\newblock \bibinfo{title}{{A deep Chandra observation of the Perseus cluster:
  shocks and ripples}}.
\newblock \emph{\bibinfo{journal}{\mnras}} \textbf{\bibinfo{volume}{344}},
  \bibinfo{pages}{L43--L47} (\bibinfo{year}{2003}).

\bibitem{2007MNRAS.381.1381S}
\bibinfo{author}{{Sanders}, J.~S.} \& \bibinfo{author}{{Fabian}, A.~C.}
\newblock \bibinfo{title}{{A deeper X-ray study of the core of the Perseus
  galaxy cluster: the power of sound waves and the distribution of metals and
  cosmic rays}}.
\newblock \emph{\bibinfo{journal}{\mnras}} \textbf{\bibinfo{volume}{381}},
  \bibinfo{pages}{1381--1399} (\bibinfo{year}{2007}).

\bibitem{2003ApJ...582L..15K}
\bibinfo{author}{{Komossa}, S.} \emph{et~al.}
\newblock \bibinfo{title}{{Discovery of a Binary Active Galactic Nucleus in the
  Ultraluminous Infrared Galaxy NGC 6240 Using Chandra}}.
\newblock \emph{\bibinfo{journal}{\apjl}} \textbf{\bibinfo{volume}{582}},
  \bibinfo{pages}{L15--L19} (\bibinfo{year}{2003}).

\bibitem{2004ApJ...600..634B}
\bibinfo{author}{{Ballo}, L.} \emph{et~al.}
\newblock \bibinfo{title}{{Arp 299: A Second Merging System with Two Active
  Nuclei?}}
\newblock \emph{\bibinfo{journal}{\apj}} \textbf{\bibinfo{volume}{600}},
  \bibinfo{pages}{634--639} (\bibinfo{year}{2004}).

\bibitem{2008MNRAS.386..105B}
\bibinfo{author}{{Bianchi}, S.}, \bibinfo{author}{{Chiaberge}, M.},
  \bibinfo{author}{{Piconcelli}, E.}, \bibinfo{author}{{Guainazzi}, M.} \&
  \bibinfo{author}{{Matt}, G.}
\newblock \bibinfo{title}{{Chandra unveils a binary active galactic nucleus in
  Mrk 463}}.
\newblock \emph{\bibinfo{journal}{\mnras}} \textbf{\bibinfo{volume}{386}},
  \bibinfo{pages}{105--110} (\bibinfo{year}{2008}).

\bibitem{2011Natur.470...66R}
\bibinfo{author}{{Reines}, A.~E.}, \bibinfo{author}{{Sivakoff}, G.~R.},
  \bibinfo{author}{{Johnson}, K.~E.} \& \bibinfo{author}{{Brogan}, C.~L.}
\newblock \bibinfo{title}{{An actively accreting massive black hole in the
  dwarf starburst galaxy Henize2-10}}.
\newblock \emph{\bibinfo{journal}{\nat}} \textbf{\bibinfo{volume}{470}},
  \bibinfo{pages}{66--68} (\bibinfo{year}{2011}).

\bibitem{2015ApJ...809L..14B}
\bibinfo{author}{{Baldassare}, V.~F.}, \bibinfo{author}{{Reines}, A.~E.},
  \bibinfo{author}{{Gallo}, E.} \& \bibinfo{author}{{Greene}, J.~E.}
\newblock \bibinfo{title}{{A {\ensuremath{\sim}}50,000
  M$_{{\ensuremath{\odot}}}$ Solar Mass Black Hole in the Nucleus of RGG 118}}.
\newblock \emph{\bibinfo{journal}{\apjl}} \textbf{\bibinfo{volume}{809}},
  \bibinfo{pages}{L14} (\bibinfo{year}{2015}).

\bibitem{2017ApJ...836...20B}
\bibinfo{author}{{Baldassare}, V.~F.}, \bibinfo{author}{{Reines}, A.~E.},
  \bibinfo{author}{{Gallo}, E.} \& \bibinfo{author}{{Greene}, J.~E.}
\newblock \bibinfo{title}{{X-ray and Ultraviolet Properties of AGNs in Nearby
  Dwarf Galaxies}}.
\newblock \emph{\bibinfo{journal}{\apj}} \textbf{\bibinfo{volume}{836}},
  \bibinfo{pages}{20} (\bibinfo{year}{2017}).

\bibitem{2000ApJ...540L..69S}
\bibinfo{author}{{Schwartz}, D.~A.} \emph{et~al.}
\newblock \bibinfo{title}{{Chandra Discovery of a 100 kiloparsec X-Ray Jet in
  PKS 0637-752}}.
\newblock \emph{\bibinfo{journal}{\apjl}} \textbf{\bibinfo{volume}{540}},
  \bibinfo{pages}{69--72} (\bibinfo{year}{2000}).

\bibitem{2002ApJ...564..683M}
\bibinfo{author}{{Marshall}, H.~L.} \emph{et~al.}
\newblock \bibinfo{title}{{A High-Resolution X-Ray Image of the Jet in M87}}.
\newblock \emph{\bibinfo{journal}{\apj}} \textbf{\bibinfo{volume}{564}},
  \bibinfo{pages}{683--687} (\bibinfo{year}{2002}).

\bibitem{2019ApJ...879....8S}
\bibinfo{author}{{Snios}, B.} \emph{et~al.}
\newblock \bibinfo{title}{{Detection of Superluminal Motion in the X-Ray Jet of
  M87}}.
\newblock \emph{\bibinfo{journal}{\apj}} \textbf{\bibinfo{volume}{879}},
  \bibinfo{pages}{8} (\bibinfo{year}{2019}).

\bibitem{2001MNRAS.321L...1C}
\bibinfo{author}{{Celotti}, A.}, \bibinfo{author}{{Ghisellini}, G.} \&
  \bibinfo{author}{{Chiaberge}, M.}
\newblock \bibinfo{title}{{Large-scale jets in active galactic nuclei:
  multiwavelength mapping}}.
\newblock \emph{\bibinfo{journal}{\mnras}} \textbf{\bibinfo{volume}{321}},
  \bibinfo{pages}{L1--L5} (\bibinfo{year}{2001}).

\bibitem{2001ApJ...551L..27M}
\bibinfo{author}{{Madau}, P.} \& \bibinfo{author}{{Rees}, M.~J.}
\newblock \bibinfo{title}{{Massive Black Holes as Population III Remnants}}.
\newblock \emph{\bibinfo{journal}{\apjl}} \textbf{\bibinfo{volume}{551}},
  \bibinfo{pages}{L27--L30} (\bibinfo{year}{2001}).

\bibitem{2006MNRAS.371.1813L}
\bibinfo{author}{{Lodato}, G.} \& \bibinfo{author}{{Natarajan}, P.}
\newblock \bibinfo{title}{{Supermassive black hole formation during the
  assembly of pre-galactic discs}}.
\newblock \emph{\bibinfo{journal}{\mnras}} \textbf{\bibinfo{volume}{371}},
  \bibinfo{pages}{1813--1823} (\bibinfo{year}{2006}).

\bibitem{2024NatAs...8..126B}
\bibinfo{author}{{Bogd{\'a}n}, {\'A}.} \emph{et~al.}
\newblock \bibinfo{title}{{Evidence for heavy-seed origin of early supermassive
  black holes from a z {\ensuremath{\approx}} 10 X-ray quasar}}.
\newblock \emph{\bibinfo{journal}{\nastro}} \textbf{\bibinfo{volume}{8}},
  \bibinfo{pages}{126--133} (\bibinfo{year}{2024}).

\bibitem{2023ApJ...955L..24G}
\bibinfo{author}{{Goulding}, A.~D.} \emph{et~al.}
\newblock \bibinfo{title}{{UNCOVER: The Growth of the First Massive Black Holes
  from JWST/NIRSpec-Spectroscopic Redshift Confirmation of an X-Ray Luminous
  AGN at z = 10.1}}.
\newblock \emph{\bibinfo{journal}{\apjl}} \textbf{\bibinfo{volume}{955}},
  \bibinfo{pages}{L24} (\bibinfo{year}{2023}).

\bibitem{2024ApJ...965L..21K}
\bibinfo{author}{{Kov{\'a}cs}, O.~E.} \emph{et~al.}
\newblock \bibinfo{title}{{A Candidate Supermassive Black Hole in a
  Gravitationally Lensed Galaxy at Z {\ensuremath{\approx}} 10}}.
\newblock \emph{\bibinfo{journal}{\apjl}} \textbf{\bibinfo{volume}{965}},
  \bibinfo{pages}{L21} (\bibinfo{year}{2024}).

\bibitem{1962PhRvL...9..439G}
\bibinfo{author}{{Giacconi}, R.}, \bibinfo{author}{{Gursky}, H.},
  \bibinfo{author}{{Paolini}, F.~R.} \& \bibinfo{author}{{Rossi}, B.~B.}
\newblock \bibinfo{title}{{Evidence for x Rays From Sources Outside the Solar
  System}}.
\newblock \emph{\bibinfo{journal}{\prl}} \textbf{\bibinfo{volume}{9}},
  \bibinfo{pages}{439--443} (\bibinfo{year}{1962}).

\bibitem{2019cxro.book....4D}
\bibinfo{author}{{Drake}, J.~J.}
\newblock \bibinfo{title}{ in \textit{The chandra x-ray observatory}} (eds
  \bibinfo{editor}{{Wilkes}, B.} \& \bibinfo{editor}{{Tucker}, W.})
  \emph{\bibinfo{booktitle}{{X-Rays from Stars and Planetary Systems}}}
  (\bibinfo{publisher}{IOP Publishing, Bristol, UK}, \bibinfo{year}{2019}).

\bibitem{2014A&A...565L...1P}
\bibinfo{author}{{Poppenhaeger}, K.} \& \bibinfo{author}{{Wolk}, S.~J.}
\newblock \bibinfo{title}{{Indications for an influence of hot Jupiters on the
  rotation and activity of their host stars}}.
\newblock \emph{\bibinfo{journal}{\aap}} \textbf{\bibinfo{volume}{565}},
  \bibinfo{pages}{L1} (\bibinfo{year}{2014}).

\bibitem{2013ApJ...773...62P}
\bibinfo{author}{{Poppenhaeger}, K.}, \bibinfo{author}{{Schmitt}, J.~H.~M.~M.}
  \& \bibinfo{author}{{Wolk}, S.~J.}
\newblock \bibinfo{title}{{Transit Observations of the Hot Jupiter HD 189733b
  at X-Ray Wavelengths}}.
\newblock \emph{\bibinfo{journal}{\apj}} \textbf{\bibinfo{volume}{773}},
  \bibinfo{pages}{62} (\bibinfo{year}{2013}).

\bibitem{2023MNRAS.524.5954I}
\bibinfo{author}{{Ili{\'c}}, N.} \emph{et~al.}
\newblock \bibinfo{title}{{The first evidence of tidally induced activity in a
  brown dwarf-M dwarf pair: a Chandra study of the NLTT 41135/41136 system}}.
\newblock \emph{\bibinfo{journal}{\mnras}} \textbf{\bibinfo{volume}{524}},
  \bibinfo{pages}{5954--5970} (\bibinfo{year}{2023}).

\bibitem{2020AJ....160..237F}
\bibinfo{author}{{France}, K.} \emph{et~al.}
\newblock \bibinfo{title}{{The High-energy Radiation Environment around a 10
  Gyr M Dwarf: Habitable at Last?}}
\newblock \emph{\bibinfo{journal}{\aj}} \textbf{\bibinfo{volume}{160}},
  \bibinfo{pages}{237} (\bibinfo{year}{2020}).

\bibitem{2024A&A...687A.237R}
\bibinfo{author}{{Rukdee}, S.} \emph{et~al.}
\newblock \bibinfo{title}{{X-ray variability of the triplet star system LTT1445
  and evaporation history of the planets around its A component}}.
\newblock \emph{\bibinfo{journal}{\aap}} \textbf{\bibinfo{volume}{687}},
  \bibinfo{pages}{A237} (\bibinfo{year}{2024}).

\bibitem{2025A&A...694A..93Z}
\bibinfo{author}{{Zhu}, E.} \& \bibinfo{author}{{Preibisch}, T.}
\newblock \bibinfo{title}{{X-ray activity of nearby G-, K-, and M-type stars
  and implications for planet habitability around M stars}}.
\newblock \emph{\bibinfo{journal}{\aap}} \textbf{\bibinfo{volume}{694}},
  \bibinfo{pages}{A93} (\bibinfo{year}{2025}).

\bibitem{2021ApJ...916...32G}
\bibinfo{author}{{Getman}, K.~V.} \& \bibinfo{author}{{Feigelson}, E.~D.}
\newblock \bibinfo{title}{{X-Ray Superflares from Pre-main-sequence Stars:
  Flare Energetics and Frequency}}.
\newblock \emph{\bibinfo{journal}{\apj}} \textbf{\bibinfo{volume}{916}},
  \bibinfo{pages}{32} (\bibinfo{year}{2021}).

\bibitem{2024ApJS..275....1B}
\bibinfo{author}{{Binder}, B.~A.} \emph{et~al.}
\newblock \bibinfo{title}{{X-Ray Emission of Nearby Low-mass and Sunlike Stars
  with Directly Imageable Habitable Zones}}.
\newblock \emph{\bibinfo{journal}{\apjs}} \textbf{\bibinfo{volume}{275}},
  \bibinfo{pages}{1} (\bibinfo{year}{2024}).

\bibitem{2013ApJ...765L...5S}
\bibinfo{author}{{Steiner}, A.~W.}, \bibinfo{author}{{Lattimer}, J.~M.} \&
  \bibinfo{author}{{Brown}, E.~F.}
\newblock \bibinfo{title}{{The Neutron Star Mass-Radius Relation and the
  Equation of State of Dense Matter}}.
\newblock \emph{\bibinfo{journal}{\apjl}} \textbf{\bibinfo{volume}{765}},
  \bibinfo{pages}{L5} (\bibinfo{year}{2013}).

\bibitem{2018MNRAS.478.2576M}
\bibinfo{author}{{Mezcua}, M.} \emph{et~al.}
\newblock \bibinfo{title}{{Intermediate-mass black holes in dwarf galaxies out
  to redshift {\ensuremath{\sim}}2.4 in the Chandra COSMOS-Legacy Survey}}.
\newblock \emph{\bibinfo{journal}{\mnras}} \textbf{\bibinfo{volume}{478}},
  \bibinfo{pages}{2576--2591} (\bibinfo{year}{2018}).

\bibitem{2020ApJ...890...59R}
\bibinfo{author}{{Reynolds}, C.~S.} \emph{et~al.}
\newblock \bibinfo{title}{{Astrophysical Limits on Very Light Axion-like
  Particles from Chandra Grating Spectroscopy of NGC 1275}}.
\newblock \emph{\bibinfo{journal}{\apj}} \textbf{\bibinfo{volume}{890}},
  \bibinfo{pages}{59} (\bibinfo{year}{2020}).

\bibitem{2007JASTP..69..179B}
\bibinfo{author}{{Bhardwaj}, A.} \emph{et~al.}
\newblock \bibinfo{title}{{First terrestrial soft X-ray auroral observation by
  the Chandra X-ray Observatory}}.
\newblock \emph{\bibinfo{journal}{\jastp}} \textbf{\bibinfo{volume}{69}},
  \bibinfo{pages}{179--187} (\bibinfo{year}{2007}).

\bibitem{2005ApJ...627L..73B}
\bibinfo{author}{{Bhardwaj}, A.} \emph{et~al.}
\newblock \bibinfo{title}{{The Discovery of Oxygen K{\ensuremath{\alpha}} X-Ray
  Emission from the Rings of Saturn}}.
\newblock \emph{\bibinfo{journal}{\apjl}} \textbf{\bibinfo{volume}{627}},
  \bibinfo{pages}{L73--L76} (\bibinfo{year}{2005}).

\bibitem{2004ApJ...607.1065M}
\bibinfo{author}{{Mori}, K.} \emph{et~al.}
\newblock \bibinfo{title}{{An X-Ray Measurement of Titan's Atmospheric Extent
  from Its Transit of the Crab Nebula}}.
\newblock \emph{\bibinfo{journal}{\apj}} \textbf{\bibinfo{volume}{607}},
  \bibinfo{pages}{1065--1069} (\bibinfo{year}{2004}).

\bibitem{2016ApJ...818..199S}
\bibinfo{author}{{Snios}, B.} \emph{et~al.}
\newblock \bibinfo{title}{{Chandra Observations of Comets C/2012 S1 (ISON) and
  C/2011 L4 (PanSTARRS)}}.
\newblock \emph{\bibinfo{journal}{\apj}} \textbf{\bibinfo{volume}{818}},
  \bibinfo{pages}{199} (\bibinfo{year}{2016}).

\end{thebibliography}


\bmhead{Acknowledgements}
PS and \'AB acknowledge support from NASA Contract NAS8-03060.

\bmhead{Author contributions}
PS, \'AB, and DP devised, together, the concept and structure of the Review.
PS led writing efforts for Sections 1, 3.1-3.4, 4, and 5. DP led writing
efforts for Section 2. \'AB led writing efforts for Sections 3.5-3.7. All
authors were responsible for final editing and review.

\bmhead{Competing interests}
The authors declare no competing interests.

\end{document}